\documentclass[reprint,onecolumn,amsmath,amssymb,aps,nofootinbib,pre]{revtex4-2}

\usepackage{graphicx} 
\usepackage[colorlinks=true,linkcolor=blue,citecolor=blue,urlcolor=blue]{hyperref}

\usepackage{bm}
\usepackage{comment}
\allowdisplaybreaks

\begin{document}
\title{Quantum tunneling Mpemba effect}
\author{Hisao Hayakawa}
\email[e-mail: ]{hisao@yukawa.kyoto-u.ac.jp}
\affiliation{Center for Gravitational Physics and Quantum Information, 
    Yukawa Institute for Theoretical Physics, 
    Kyoto University, 
    Kitashirakawa-Oiwakecho, Sakyo-ku, Kyoto 606-8502, Japan}
\author{Satoshi Takada}
\email[e-mail: ]{takada@go.tuat.ac.jp}
\affiliation{Department of Mechanical Systems Engineering and 
    Institute of Engineering, 
    Tokyo University of Agriculture and Technology, 
    2-24-16 Naka-cho, Koganei, Tokyo 184-8588, Japan}

\date{\today}

\begin{abstract}
We investigate a quantum tunneling Mpemba effect for a particle in a continuous one-dimensional symmetric double-well potential subject to irreversible boundary loss. 
The dissipation is described by a complex absorbing potential (CAP) and the corresponding conditional no-jump evolution. 
Using a biorthogonal spectral representation, we derive the exact non-Hermitian decomposition of the subnormalized density matrix and analyze the corresponding relaxation dynamics. 
We show that the off-diagonal coefficients exhibit a parity selection rule, leaving same-parity interference terms in general. 
To isolate the interplay between thermal spectral preparation and mode-dependent decay rates, we introduce a diagonal spectral relaxation measure $S(t,T_i)$. 
For a quartic double well, numerical calculations reveal finite-time crossings of $S(t,T_i)$ prepared at different initial temperatures, as well as a non-monotonic temperature dependence of selected diagonal spectral weights. 
These features are explained by the enhanced thermal population of spatially extended, rapidly leaking excited states. 
In contrast, the trace distance of the normalized conditional state to the asymptotic quasi-stationary state exhibits no crossing, demonstrating that the anomalous ordering is specifically a spectral Mpemba effect rather than a universal acceleration of the complete conditional state. 
Our continuous-space formulation provides a real-space perspective on anomalous quantum relaxation without invoking universal nodal or topological theorems.  
\end{abstract}

\maketitle
\section{Introduction}
The Mpemba effect is a counterintuitive relaxation phenomenon in which an initially hotter system cools or relaxes to the ambient environment faster than an initially colder one. 
First reported in modern science by Mpemba and Osborne within the context of freezing water mixtures~\cite{mpemba1969cool}, the physical reproducibility of the effect originally sparked deep debates regarding the influence of uncontrolled external variables~\cite{burridge2016questioning, katz2017reply}. 
Over the past decade, however, the phenomenon has been re-established as a reproducible effect within the framework of modern non-equilibrium statistical mechanics, as thoroughly reviewed in recent extensive literature~\cite{Teza25, Bechhoefer2021}. 
Modern breakthroughs have successfully identified and analyzed the effect across a vast array of classical systems, including Markovian stochastic processes~\cite{Lu17, Klich19, Busiello21,Pemartin23}, non-Markovian dynamics~\cite{Yang20}, a spin-glass~\cite{Baity-Jesi19}, quench dynamics of Ising model~\cite{S_Chatterjee24}, granular fluids and gases~\cite{Santos17, Torrente19, Biswas20, Biswas21, Mompo20, Megias22b, Patron23, Santos20, Patron21, Takada21a, Megias22a, Biswas23,Gonzalez2021}, carbon nanotube resonators~\cite{Greaney11}, magnetic systems~\cite{Gonzalez21}, systems experiencing phase transitions~\cite{Holtzman22}, a heat engine~\cite{Lin22}, a delay differential equation~\cite{Santos24}, and active matters~\cite{Pal24,Antonov26}.

Crucially, the classical Mpemba effect has been precisely quantified in high-precision desktop experiments. 
Outstanding examples include the cooling and heating dynamics of optical-tweezer-trapped colloidal particles~\cite{kumar2020, kumar2021, Kumar22, Malhotra24}, as well as energy-dissipating Langevin setups \cite{Tian2025, Biswas23a, Biswas_thesis, chetrite2021metastable, walker2021anomalous, Deguenther22}. 
See also general arguments for the conditions to observe classical Mpemba effects~\cite{Ohga2024,Vu2025}.
A central question that has persisted throughout these studies is identifying the fundamental architectural features of the potential energy landscape required to trigger such anomalous relaxation. 
Recently, a major paradigm shift has occurred: while earlier theories pinned the effect exclusively on the presence of multiple internal metastable states, a series of exact investigations has established that the phenomenon can be observed in single-well or monotonic potentials, provided that boundary configurations, such as hard or soft walls, are properly accounted for~\cite{Yue26, Yue_long26,Hayakawa2026}. 
This breakthrough highlights that the classical Mpemba effect ``likes to hit a wall''~\cite{Yue26,Yue_long26}, making boundary constraints a primary universal driver for anomalous cooling.

Paralleling these classical milestones, the study of the Mpemba effect has rapidly expanded into the quantum domain, transforming into a vibrant frontier of quantum thermodynamics~\cite{Ares25_review, Sagawa}. 
Advanced theoretical and numerical frameworks have characterized the quantum Mpemba effect through various mechanisms, such as Gorini–Kossakowski–Sudarshan–Lindblad (GKSL) master equation dynamics~\cite{Nava19, Carollo21, Manikandan21, Ivander23,Moroder24}, dynamical symmetry restoration and entanglement asymmetry in spin chains or free fermions~\cite{Ares23, Joshi2024, Rylands2024, Chalas24, Ares24, Yamashika2024, Liu2024, Turkeshi2025, Yamashika2026}, a resonance state~\cite{Yamashika26}, non-Markovian open systems~\cite{Yang20,Strachan25, Yang22b, Wang2024, Nava2024}, and quantum dots coupled to reservoirs~\cite{Chatterjee_2023, Chatterjee24}. 
Furthermore, the quantum effect has achieved definitive experimental verification across platforms ranging from single trapped-ion qubits~\cite{Shapira2024} to dissipative non-Hermitian acoustic or photonic setups~\cite{Longhi2024, Zhang2025}, as well as through innovative temporary quantum reset strategies~\cite{Bao2025, Beato2026}. 
More broadly, ultracold atoms in optical lattices with controlled one-body loss provide an experimentally established platform for realizing non-Hermitian many-body dynamics and monitoring the associated loss processes~\cite{Takasu2020}.

Yet, despite this profound progress, a conspicuous gap remains between the classical and quantum pictures. 
While the classical framework has explicitly isolated the spatial connection between boundary constraints and anomalous particle evaporation in real space~\cite{Yue26,Yue_long26}, its quantum-mechanical counterpart, incorporating wave-function coherence, spatial barriers, and non-Hermitian boundary leakage, has not yet been thoroughly formulated in a continuous coordinate space. 
Most existing quantum studies focus on abstract discrete spin configurations or abstract state-space quenches rather than physical tunneling across continuous potentials. 

In this paper, we investigate a continuous-space realization of the quantum tunneling Mpemba effect.
We analyze a single particle trapped within a continuous one-dimensional symmetric double-well potential that is explicitly opened to the environment via absorbing boundary sinks at $x=\pm L$.
The boundary leakage is governed by a complex non-Hermitian effective Hamiltonian with complex absorbing potentials, which we discretize directly in coordinate space.
By employing a biorthogonal spectral decomposition, we derive the exact mode representation of the subnormalized conditional density matrix while rigorously retaining same-parity interference terms. 
To isolate the interplay between thermal spectral preparation and mode-dependent decay rates, we introduce a diagonal spectral relaxation measure $S(t,T_i)$. 
For a quartic double well, numerical calculations demonstrate finite-time crossings of $S(t,T_i)$ prepared at different initial temperatures, as well as a non-monotonic initial-temperature dependence of selected diagonal spectral weights. 
Furthermore, by evaluating the trace distance of the normalized conditional state to the asymptotic quasi-stationary state, we confirm that no crossing occurs in the full state space, establishing that the anomalous ordering is specifically a spectral Mpemba effect. 
Our continuous formulation successfully unifies the real-space intuition of classical boundary-driven dissipation with the fundamental principles of open quantum dynamics.

The contents of this paper are as follows.
In Sec.~\ref{Sec:II}, we formulate the model used in this paper based on non-Hermitian quantum mechanics with a complex absorbing potential near the boundary.
In Sec.~\ref{Sec:S(t,T_i)}, we establish the full non-Hermitian spectral decomposition (biorthogonal bases, eigenfrequencies, parity selection rules, density matrix expansion, and the definition of $S(t,T_i)$). 
In Sec.~\ref{Sec:confirmation}, we numerically verify the theoretical results proposed in Sec.~\ref{Sec:S(t,T_i)}, including the evaluation of spectral measure crossings and conditional trace distances.
In Sec.~\ref{Sec:Discussion}, we discuss the underlying physical mechanisms and spatial mode overlaps.
In Sec.~\ref{Sec:Conclusion}, we present the concluding remarks.
In the appendices, we show some technical details used in the main text.

\section{NON-HERMITIAN DENSITY MATRIX DYNAMICS UNDER BOUNDARY DISSIPATION}\label{Sec:II}
We begin our analysis of the quantum tunneling Mpemba effect by considering the open system dynamics described by the GKSL master equation. 
For a density matrix $\rho$, the general form is:
\begin{equation}\label{GKSL}
    \partial_t \hat\rho = -\frac{i}{\hbar} [\hat{H}_0, \hat\rho] + \sum_k \gamma_k \left( \hat{L}_k \hat{\rho} \hat{L}_k^\dagger - \frac{1}{2} \{ \hat{L}_k^\dagger \hat{L}_k, \hat\rho \} \right),
\end{equation}
where $\hat{H}_0:=p^2/(2m)+V(x)$ is the Hermitian Hamiltonian of the system, $\hat{L}_k$ denotes the Lindblad jump operators representing specific environmental dissipation channels, and $\{\cdot,\cdot\}$ is the anti-commutator. 
In this study, $V(x)$ is a symmetric double-well potential, whose explicit form is specified in Sec.~\ref{Sec:confirmation}.

To formulate the continuous quantum tunneling Mpemba effect under an open-system architecture, we introduce a smoothly varying Complex Absorbing Potential (CAP) into the physical landscape rather than enforcing terminal point-like boundary constraints~\cite{Riss1993,Muga2004}. 
Non-equilibrium relaxation and particle loss dynamics of a quantum particle of mass $m$ trapped within a smooth, reflection-symmetric double-well potential $V(x) = V(-x)$ are governed by the time-dependent Schr\"{o}dinger equation for the wave function $\psi(x,t):=\langle x|\psi(t)\rangle$ under a non-Hermitian effective Hamiltonian $\hat{H}_{\text{eff}}: = \hat{H}_0 - i\hat{W}$:
\begin{align}\label{CAP_Schrodinger}
    i\hbar \frac{\partial}{\partial t}\psi(x,t) &= {H}_\mathrm{eff} \psi(x,t),
\end{align}
where 
\begin{align}\label{def:H_eff}
    H_\mathrm{eff}:=H_0(x)-i W(x)
    = -\frac{\hbar^2}{2m}\frac{\partial^2}{\partial x^2} 
    + V(x) - iW(x) 
\end{align}
with the mass $m$ of a tracer particle.
Here, the continuous spatial absorption profile $W(x) \ge 0$ acts as an isotropic dissipative buffer. 
We note that $|\psi(t)\rangle$ satisfies $i\hbar \frac{\partial}{\partial t}|\psi(t)\rangle = \hat{H}_{\text{eff}} |\psi(t)\rangle$ with the relation $\hat{H}_\mathrm{eff}=\int_{-L}^L dx|x\rangle \langle x| H_\mathrm{eff}$.
To ensure that the coherent quantum mechanics and tunneling splits remain unperturbed within the inner core of the bistable landscape, the CAP is set to zero inside the wells and ramps up smoothly only as the particle approaches the outer system boundaries at $x = \pm L$. 
A common numerical and analytical choice is the power-law profile confined to boundary layers of width $x_{\text{abs}}$:
\begin{equation}\label{W(x)}
    W(x) = \begin{cases} 
    0 & |x| \le L - x_{\text{abs}}, \\
    \eta \left( |x| - (L - x_{\text{abs}}) \right)^n & L - x_{\text{abs}} < |x| \le L,
    \end{cases}
\end{equation}
where $\eta > 0$ regulates the continuous absorption strength and $n \in \mathbb{N}$ controls the smoothness of the dissipative onset to minimize unphysical quantum reflections back into the central wells.
The parameter $\eta$ controls the effective loss strength in the CAP model. 
The present formulation does not rely on a microscopic weak- or strong-coupling expansion of an environmental bath.
The weak-CAP assumption introduced later is used only as an auxiliary perturbative argument and is not required for the exact spectral representation.

To provide a firm open-quantum-systems foundation for this formulation, we establish the explicit mathematical mapping between the phenomenological CAP framework and the underlying Markovian master equation. 
The full, trace-preserving temporal evolution of the system's density matrix $\hat{\rho}(t)$ interacting with an external environment is fundamentally described by the GKSL master equation \eqref{GKSL}. 
By regrouping the coherent unitary terms and the dissipative anti-commutator, Eq.~\eqref{GKSL} can be identically rewritten as:
\begin{equation}\label{Eq5}
    \frac{\partial \hat{\rho}}{\partial t} =\mathcal{L}(\hat{\rho}), 
    \quad 
    \mathcal{L}(\hat{\rho}):=-\frac{i}{\hbar} \left( \hat{H}_{\text{eff}} \hat{\rho} - \hat{\rho} \hat{H}_{\text{eff}}^\dagger \right) + \sum_{\mu} \hat{L}_\mu \hat{\rho} \hat{L}_\mu^\dagger,
\end{equation}
where the non-Hermitian effective Hamiltonian is formally defined as 
\begin{align}\label{H_{eff}}
    \hat{H}_{\text{eff}}:= \hat{H}_0 - \frac{i\hbar}{2} \sum_{\mu} \hat{L}_\mu^\dagger \hat{L}_\mu .   
\end{align}

The final term in Eq.~\eqref{Eq5}, $\sum_{\mu} \hat{L}_\mu \hat{\rho} \hat{L}_\mu^\dagger$, acts as a recycling term that feeds probability back into the system's state space after a quantum jump occurs. 
However, in an evaporation or leakage experiment where particles are permanently absorbed upon reaching the boundaries $x = \pm L$, the recycled states correspond exclusively to particles that have already left the core double well and entered the vacuum reservoir. 
Physically, tracking the survival probability within the wells amounts to conditioning the master equation on the sub-ensemble where \textit{zero quantum jumps have occurred} up to time $t$~\cite{Dalibard1992, Plenio1998, Carollo21}. 
Setting the recycling term to zero yields the conditional non-unitary dynamics of the surviving state:
\begin{equation}\label{Eq6}
    \frac{\partial \hat{\rho}_c}{\partial t} = -\frac{i}{\hbar} \left( \hat{H}_{\text{eff}} \hat{\rho}_c - \hat{\rho}_c \hat{H}_{\text{eff}}^\dagger \right),
\end{equation}
which, for an initially pure state $\hat{\rho}_c(0) = |\psi(0)\rangle\langle\psi(0)|$, is identical to the non-Hermitian Schr\"{o}dinger equation for $\hat{H}_\mathrm{eff}$.
While Eq.~\eqref{Eq6} preserves its structural form for the mixed initial thermal state $\hat{\rho}(0) = \sum_m P_m(T_i)|\varphi_m\rangle\langle\varphi_m|$, its equivalence to the state-vector non-Hermitian Schr\"{o}dinger equation holds independently for each underlying pure statistical sub-ensemble component $|\varphi_m(t)\rangle = e^{-i\hat{H}_{\text{eff}}t/\hbar}|\varphi_m\rangle$. 
The mixed thermal distributions are subsequently mapped onto the spectral amplitudes, as will be discussed later.
Thus, the time evolution of the density matrix can be expressed as
\begin{align}\label{rho_c(t)}
    \hat{\rho}_c(t) = e^{-i \hat{H}_{\text{eff}} t/\hbar} \hat{\rho}(0) e^{i \hat{H}_{\text{eff}}^\dagger t/\hbar}.
\end{align}

To explicitly connect this operator formulation to our continuous coordinate space, we model the boundary loss by localized jump operators $\hat{L}(x)$ that transfer a particle in the dissipation zone to an orthogonal vacuum sector:
\begin{equation}\label{Eq9}
    \hat{L}(x) = \sqrt{\frac{2}{\hbar} W(x)} |\mathrm{vac}\rangle\langle x|,
\end{equation}
where $W(x)$ is the continuous spatial absorption profile introduced in Eq.~\eqref{W(x)}~\cite{Riss1993, Muga2004, Nava2024}. 

Here, $|\mathrm{vac}\rangle$ introduced in Eq.~\eqref{Eq9} denotes
the normalized zero-particle state in the enlarged Hilbert space
\begin{equation}
    \mathcal{H}_{\mathrm{full}}
    =
    \mathcal{H}_{\mathrm{particle}}
    \oplus
    \operatorname{span}\{|\mathrm{vac}\rangle\},
    \qquad
    \langle\mathrm{vac}|\psi\rangle=0
    \quad
    (|\psi\rangle\in\mathcal{H}_{\mathrm{particle}}).
\end{equation}
It represents the absence of the particle after a loss event and should not be confused with the one-particle ground state of $\hat H_0$. 
Extending the effective Hamiltonian to the vacuum sector, we write
\begin{equation}
    \hat H_{\mathrm{eff}}^{\mathrm{full}}
    =
    (\hat H_0-i\hat W)\oplus E_{\mathrm{vac}},
    \qquad
    \hat H_{\mathrm{eff}}^{\mathrm{full}}
    |\mathrm{vac}\rangle
    =
    E_{\mathrm{vac}}|\mathrm{vac}\rangle,
\end{equation}
where the real vacuum energy $E_{\mathrm{vac}}$ may be set to zero.
Thus, $|\mathrm{vac}\rangle$ belongs to the zero-particle sector and is not an eigenstate of the one-particle operator $\hat H_{\mathrm{eff}}=\hat H_0-i\hat W$.

The corresponding vacuum density operator is defined as
\begin{equation}
    \hat{\rho}_{\mathrm{vac}}
    :=
    |\mathrm{vac}\rangle\langle\mathrm{vac}|.
\end{equation}
Because the loss jump operators annihilate the vacuum,
\begin{equation}
    \hat{L}(x)|\mathrm{vac}\rangle=0,
\end{equation}
and because $E_{\mathrm{vac}}$ is real, the vacuum density operator is a stationary zero-eigenvalue eigenoperator of the full Liouvillian:
\begin{equation}
    \mathcal{L}
    \left(\hat{\rho}_{\mathrm{vac}}\right)
    =0.
\end{equation}
Therefore, $\hat{\rho}_{\mathrm{vac}}$ is the absorbing stationary state of the trace-preserving dynamics in the enlarged particle-plus-vacuum Hilbert space.

Evaluating the anti-commutator sum over this continuous position basis transforms the sum into a spatial integral over the domain:
\begin{equation}
    \frac{\hbar}{2} \sum_{\mu} \hat{L}_\mu^\dagger \hat{L}_\mu \longrightarrow \frac{\hbar}{2} \int_{-L}^{L} \hat{L}^\dagger(x) \hat{L}(x) \, dx = \int_{-L}^{L} W(x) |x\rangle\langle x| \, dx = \hat{W}.
\end{equation}
Substituting this result directly into the definition of $\hat{H}_{\text{eff}}$ in Eq.~\eqref{H_{eff}} yields $\hat{H}_{\text{eff}} = \hat{H}_0 - i\hat{W}$, which exactly reproduces the coordinate-space potential $V(x) - iW(x)$ given in Eq.~\eqref{def:H_eff}. 
Thus the CAP is the no-jump generator associated with irreversible loss to the vacuum sector.  
A projector-valued jump operator proportional to $|x\rangle\langle x|$ would instead describe local measurement or dephasing and would not represent particle removal.

Since $H_{\text{eff}}$ contains the tunneling information in its real part and the boundary loss in its imaginary part, the decay of the trace $S(t,T_i)$ provides a direct measure of the relaxation toward the empty state. 
This allows us to compare the relaxation rates of different initial thermal states $\hat{\rho}(T_\mathrm{ini})$ to identify the existence of the quantum tunneling Mpemba effect.

\section{BIORTHOGONAL SPECTRAL DECOMPOSITION AND RELAXATION DYNAMICS}\label{Sec:S(t,T_i)}
To analyze the relaxation dynamics of the density matrix $\hat{\rho}(t)$, we utilize the spectral properties of the effective Hamiltonian $\hat{H}_{\text{eff}}$. 
In the no-jump limit, the evolution Eq.~\eqref{Eq6} can be decomposed by considering the evolution of individual state vectors via the non-Hermitian Schr\"{o}dinger equation.

\subsection{Spectral Decomposition and Biorthogonality}
Let $\{ |\phi_n \rangle \}$ be the right-eigenstates of $H_{\text{eff}}$ with complex eigenvalues $\lambda_n$:
\begin{equation}\label{eigenequation}
    \hat{H}_{\text{eff}} |\phi_n \rangle = \lambda_n |\phi_n \rangle, \quad \lambda_n = E_n - i \frac{\Gamma_n}{2},
\end{equation}
where $E_n$ is the energy (the eigenvalue of $\hat{H}_0$) and $\Gamma_n \ge 0$ is the decay rate. The adjoint equation is 
\begin{equation}\label{H_eff^d}
 \hat{H}_{\text{eff}}^\dagger |\chi_n \rangle = \lambda_n^* |\chi_n \rangle,   
\end{equation}
where the left-eigenstates $\{ \langle \chi_n | \}$ form a biorthogonal set with $\{ |\phi_m \rangle \}$ such that $\langle \chi_n | \phi_m \rangle = \delta_{nm}$.

Because $\hat H_{\mathrm{eff}}$ describes an open system with absorbing regions, its eigenvalues are generally complex.
The imaginary parts determine the characteristic decay rates of the resonance modes.  
In the parameter regime considered below, the relevant modes satisfy
\begin{equation}
    \Gamma_k>0,
\end{equation}
so that every individual resonance contribution is progressively suppressed during the evolution.

Since the absorbing layers are introduced near the edges of the finite computational domain, we supplement Eq.~\eqref{def:H_eff}
with hard-wall Dirichlet boundary conditions at the terminal boundaries,
\begin{equation}
    \phi_k(\pm L)=0.
    \label{Dirichlet}
\end{equation}
The resulting resonance spectrum is discrete.
The right eigenstates are not, in general, orthogonal under the ordinary Hilbert-space inner product and must therefore be treated using the biorthogonal structure of the non-Hermitian problem.

For the complex-symmetric coordinate representation used in the present work, the left and right eigenfunctions are related by
\begin{equation}
    \chi_k^*(x)=\phi_k(x),
\end{equation}
and we use the normalization
\begin{equation}
    \langle\chi_k|\phi_j\rangle
    =
    \int_{-L}^{L}
    dx\,\phi_k(x)\phi_j(x)
    =
    \delta_{kj}.
    \label{eq:c_norm_main}
\end{equation}
This normalization convention is used throughout the spectral
analysis below.

\subsection{Initial Thermal Distribution and Diagonal Spectral Weights}
\label{sec:initial_thermal}
We assume that the system is initially prepared in thermal equilibrium at temperature $T_i$ with respect to the Hermitian Hamiltonian $\hat H_0$.
The initial density operator is
\begin{equation}
    \hat\rho_i(T_i)
    = \sum_n P_n(T_i) |n\rangle\langle n|,
    \qquad
    P_n(T_i)
    = \frac{e^{-E_n/(k_BT_i)}}{Z(T_i)},
    \label{eq:rho_initial_main}
\end{equation}
where
\begin{equation}
    Z(T_i)
    = \sum_n e^{-E_n/(k_BT_i)}
\end{equation}
is the partition function.
The closed-system eigenstates $|n\rangle$ satisfy
\begin{equation}
    \hat H_0|n\rangle
    =
    E_n|n\rangle.
    \label{eigen_H_0}
\end{equation}

To characterize how the initial thermal state is projected onto the resonance modes of the open system, we introduce the diagonal spectral coefficients
\begin{equation}
    a_k(T_i)
    :=
    \langle\chi_k|
    \hat\rho_i(T_i)
    |\chi_k\rangle.
    \label{eq:ak_def_main}
\end{equation}
Using Eq.~\eqref{eq:rho_initial_main}, this coefficient can be written as
\begin{equation}
    a_k(T_i)
    =
    \sum_n
    P_n(T_i)
    |\langle\chi_k|n\rangle|^2
    =
    \frac{1}{Z(T_i)}
    \sum_n
    e^{-E_n/(k_BT_i)}
    |O_{kn}|^2,
    \label{Eq11}
\end{equation}
where
\begin{equation}
    O_{kn}
    :=
    \langle\chi_k|n\rangle.
    \label{eq:Okn_main}
\end{equation}

The coefficient $a_k(T_i)$ characterizes the spectral weight with which the initial thermal state probes the $k$-th resonance mode.
It should not be interpreted as an ordinary mode population, since the left eigenstates of a non-Hermitian Hamiltonian do not form an orthonormal basis under the ordinary Hilbert-space inner product.
Rather, $a_k(T_i)$ provides a resonance-resolved measure of the thermal spectral preparation.

Since the resonance spectrum and the decay rates are fixed once $\hat H_{\rm eff}$ is specified, the temperature dependence of $a_k(T_i)$ arises entirely from the Gibbs weights $P_n(T_i)$.
Different initial temperatures therefore prepare different distributions of spectral weight over the same set of resonance channels.  
The subsequent relaxation associated with these spectral weights will be discussed in Sec.~\ref{sec:spectral_crossing}.

\subsection{Diagonal and Off-Diagonal Spectral Coefficients}
\label{sec:cross_modes}
The biorthogonal representation of the initial thermal state generally involves not only the diagonal coefficients $a_k(T_i)$ introduced above, but also the off-diagonal coefficients
\begin{equation}
    a_{k,j}(T_i)
    :=
    \langle\chi_k|
    \hat\rho_i(T_i)
    |\chi_j\rangle
    =
    \frac{1}{Z(T_i)}
    \sum_n
    e^{-E_n/(k_BT_i)}
    O_{kn}O_{jn}^*.
    \label{eq:akj_main}
\end{equation}
The diagonal coefficients in Eq.~\eqref{eq:ak_def_main} are recovered as
\begin{equation}
    a_k(T_i)=a_{k,k}(T_i).
\end{equation}

The off-diagonal coefficients describe cross-mode components of the thermal state in the biorthogonal resonance representation.
They are generally complex and need not vanish in a non-Hermitian system.

For the symmetric potential and CAP considered here, both the closed eigenstates $|n\rangle$ and the resonance modes have definite parity.
Consequently,
\begin{equation}
    a_{k,j}(T_i)=0
    \qquad
    \text{if $k$ and $j$ have opposite parity}.
    \label{eq:akj_parity}
\end{equation}
In contrast, off-diagonal coefficients between modes belonging to the same parity sector are generally nonzero.
This parity selection rule produces the characteristic checkerboard structure of $|a_{k,j}(T_i)|$ observed numerically below.

The distinction between diagonal and off-diagonal coefficients becomes explicit in the full non-Hermitian evolution of the conditional density operator.  
Using the biorthogonal spectral decomposition, we obtain
\begin{equation}
    \hat\rho_c(t,T_i)
    =
    \sum_{k,j}
    a_{k,j}(T_i)
    \exp\left[
        -\frac{i}{\hbar}
        (\lambda_k-\lambda_j^*)t
    \right]
    |\phi_k\rangle\langle\phi_j|.
    \label{eq:rho_full_spectral_main}
\end{equation}
The diagonal terms $k=j$ contain the individual resonance decay factors, whereas the off-diagonal terms $k\neq j$ additionally contain relative phase factors between distinct resonance energies.
Therefore, the off-diagonal coefficients are relevant to the complete conditional density-matrix dynamics.

In the following subsection, we introduce a diagonal spectral relaxation measure constructed from $a_{k,k}(T_i)$ alone.
This does not assume that the off-diagonal coefficients vanish.
Rather, the diagonal measure is introduced deliberately to isolate the competition between thermal spectral preparation and the mode-dependent decay rates.
The off-diagonal coefficients will be examined separately in the numerical analysis and will also enter the trace-distance analysis of the full conditional state.

\subsection{Spectral Relaxation and Mpemba Crossings}
\label{sec:spectral_crossing}
To characterize the relaxation associated specifically with the diagonal resonance spectrum, we define the spectral relaxation measure
\begin{equation}
     S(t,T_i)
    :=
    \sum_k
    a_k(T_i)
    \exp\left(
        -\frac{\Gamma_k t}{\hbar}
    \right),
    \qquad
    a_k(T_i)=a_{k,k}(T_i).
    \label{eq:Stilde}
\end{equation}
Thus, each resonance channel contributes through the product of its initial spectral weight $a_k(T_i)$ and its decay factor
$\exp(-\Gamma_k t/\hbar)$.
The quantity $S(t,T_i)$ therefore measures the decay-weighted resonance content of the initial thermal state.

An equivalent operator representation is
\begin{equation}
    S(t,T_i)
    =
    \operatorname{Tr}
    \left[
        \hat\rho_i(T_i)\hat A(t)
    \right],
    \label{eq:Stilde_operator}
\end{equation}
where
\begin{equation}
    \hat A(t)
    :=
    \sum_k
    \exp\left(
        -\frac{\Gamma_k t}{\hbar}
    \right)
    |\chi_k\rangle\langle\chi_k|.
    \label{eq:A_operator}
\end{equation}
The operator $\hat A(t)$ acts as a time-dependent spectral filter: resonances with large decay rates are progressively suppressed, whereas long-lived resonances retain their spectral weight for a longer time.

As discussed in Sec.~\ref{sec:cross_modes}, $S(t,T_i)$ intentionally contains only the diagonal spectral coefficients.  
It is therefore distinct from the full conditional density-matrix evolution in Eq.~\eqref{eq:rho_full_spectral_main}, which also contains off-diagonal cross-mode contributions.
This construction allows us to isolate the competition between the temperature-dependent resonance weights and the mode-dependent decay rates.

Since
\begin{equation}
    a_k(T_i)\geq0,
    \qquad
    \Gamma_k\geq0,
\end{equation}
the spectral relaxation measure is nonnegative,
\begin{equation}
     S(t,T_i)\geq0,
\end{equation}
and monotonically nonincreasing:
\begin{equation}
    \frac{\partial S(t,T_i)}{\partial t}
    =
    -\frac{1}{\hbar}
    \sum_k
    \Gamma_k
    a_k(T_i)
    \exp\left(
        -\frac{\Gamma_k t}{\hbar}
    \right)
    \leq0.
    \label{eq:Stilde_monotonic}
\end{equation}
More generally,
\begin{equation}
    (-1)^p
    \frac{\partial^p S(t,T_i)}{\partial t^p}
    =
    \frac{1}{\hbar^p}
    \sum_k
    \Gamma_k^p
    a_k(T_i)
    \exp\left(
        -\frac{\Gamma_k t}{\hbar}
    \right)
    \geq0,
    \qquad
    p=0,1,2,\ldots.
    \label{eq:Stilde_complete_monotonicity}
\end{equation}
Thus, $S(t,T_i)$ is a completely monotone relaxation function for each fixed initial temperature.

The same structure can be expressed in terms of the temperature-dependent spectral distribution over the decay rates,
\begin{equation}
    \mathcal A_{T_i}(\Gamma)
    :=
    \sum_k
    a_k(T_i)
    \delta(\Gamma-\Gamma_k).
    \label{eq:spectral_decay_distribution}
\end{equation}
Then Eq.~\eqref{eq:Stilde} becomes
\begin{equation}
     S(t,T_i)
    =
    \int_0^\infty
    d\Gamma\,
    \mathcal A_{T_i}(\Gamma)
    \exp\left(
        -\frac{\Gamma t}{\hbar}
    \right).
    \label{eq:Stilde_Laplace}
\end{equation}
Hence, the initial temperature determines the distribution of spectral weight among resonance modes with different decay rates, whereas the subsequent evolution filters this distribution according to $\Gamma$.

To identify a Mpemba crossing, we consider two initial temperatures $T_h>T_c$ and define
\begin{equation}
    \Delta S(t)
    :=
    S(t,T_h)
    -
    S(t,T_c).
    \label{eq:Delta_Stilde}
\end{equation}
An interior crossing time $t_\times>0$ satisfies
\begin{equation}
    \Delta S(t_\times)=0,
    \qquad
     S(t_\times,T_h)
    =
     S(t_\times,T_c),
    \label{eq:Stilde_crossing_condition}
\end{equation}
together with a change in the sign of $\Delta S(t)$ across $t_\times$.
For example, if
\begin{equation}
    \Delta S(t_1)>0,
    \qquad
    \Delta S(t_2)<0,
    \qquad
    0\leq t_1<t_\times<t_2,
    \label{eq:Stilde_Mpemba_sign}
\end{equation}
the ordering of the two thermal preparations is reversed during the spectral relaxation.
We refer to this finite-time inversion as a \emph{spectral Mpemba crossing}.

Although each $ S(t,T_i)$ is monotonically decreasing, their difference $\Delta S(t)$ need not be monotonic.
A crossing can therefore arise when different initial temperatures produce different distributions $a_k(T_i)$ over the common decay spectrum $\Gamma_k$.
In particular, a non-monotonic temperature dependence of one or more spectral coefficients can transfer relative spectral weight between rapidly and slowly decaying resonance channels and thereby reverse the ordering of $S(t,T_i)$.

The mechanism of the spectral Mpemba effect can thus be summarized as $\text{temperature-dependent resonance weights}\;+\;\text{mode-dependent decay rates}\;\Longrightarrow\;\text{spectral Mpemba crossing}$.
While $S(t, T_i)$ isolates the diagonal spectral relaxation dynamics, a complete description of the open quantum state requires tracking the normalized conditional density matrix $\hat{\rho}_{\text{cond}}(t): = \hat{\rho}_c(t) / \operatorname{Tr}[\hat{\rho}_c(t)]$, which will be introduced in Sec. \ref{sec:trace_distance}. 
In Sec.~\ref{Sec:confirmation}, we complement our analysis of $S(t, T_i)$ by evaluating the trace distance $D_{\text{QS}}(t, T_i)$ of $\hat{\rho}_{\text{cond}}(t)$ relative to the asymptotic quasi-stationary state $\hat{\rho}_{\text{QS}}$.

\subsection{Existence of Non-Monotonic Peaks in $a_k(T_i)$ for $k\ne 0$}
\label{SecIIID}
To characterize the temperature dependence of the individual resonance channels entering the spectral relaxation measure $S(t,T_i)$, we introduce the diagonal spectral coefficient
\begin{equation}
    a_k(T_i)
    :=
    a_{k,k}(T_i)
    =
    \frac{1}{Z(T_i)}
    \sum_{m\in\mathcal{P}_k}
    e^{-E_m/(k_BT_i)}
    |O_{km}|^2,
    \qquad
    O_{km}:=\langle\chi_k|m\rangle,
    \label{Eq18}
\end{equation}
where $\mathcal{P}_k$ denotes the set of closed-system eigenstates having the same parity as the $k$-th resonance mode.
For the complex-symmetric coordinate representation,
\begin{equation}
    O_{km}
    =
    \int_{-L}^{L}
    \phi_k(x)\varphi_m(x)\,dx,
    \qquad
    \varphi_m(x):=\langle x|m\rangle.
    \label{eq:Okm_main}
\end{equation}
Reflection symmetry implies
\begin{equation}
    O_{km}=0
\end{equation}
when the $k$-th resonance mode and the closed state $|m\rangle$ have opposite parity.

The spectral relaxation measure introduced in this work is
\begin{equation}
    S(t,T_i)
    =
    \sum_k
    a_k(T_i)
    e^{-\Gamma_k t/\hbar}.
    \label{eq:Stilde_main}
\end{equation}
Thus, the contribution of the $k$-th resonance channel is
\begin{equation}
    S_k(t,T_i)
    =
    a_k(T_i)e^{-\Gamma_k t/\hbar}.
    \label{eq:Sk_diag}
\end{equation}
Since $\Gamma_k$ is independent of $T_i$, the temperature dependence of each individual channel is entirely determined by $a_k(T_i)$.
A non-monotonic dependence of $a_k(T_i)$ therefore produces a non-monotonic thermal activation of the corresponding resonance channel.

We use the complex-symmetric normalization
\begin{equation}
    \langle\chi_n|\phi_m\rangle
    =
    \int_{-L}^{L}dx\,\phi_n(x)\phi_m(x)
    =
    \delta_{nm}.
    \label{eq:c_norm_main2}
\end{equation}
This fixes the normalization convention used below.
Although the numerical value of $a_k$ depends on the normalization of the biorthogonal eigenvectors, the temperature at which $a_k(T_i)$ for $k\ne 0$ reaches a maximum is unchanged by a temperature-independent rescaling of the $k$-th mode.

\subsubsection{Low-temperature limit}
Let
\begin{equation}
    \Delta_0:=E_1-E_0>0
\end{equation}
denote the tunneling splitting of the closed Hamiltonian.
As $T_i\to0^+$,
\begin{equation}
    Z(T_i)
    =
    e^{-E_0/(k_BT_i)}
    \left[
    1+e^{-\Delta_0/(k_BT_i)}
    +\mathcal O\!\left(
    e^{-(E_2-E_0)/(k_BT_i)}
    \right)
    \right].
\end{equation}
It follows directly from Eq.~\eqref{Eq18} that
\begin{equation}
    \lim_{T_i\to0^+}a_k(T_i)
    =
    |O_{k0}|^2.
    \label{eq:ak_lowT}
\end{equation}

For an odd resonance mode, reflection symmetry gives
\begin{equation}
    O_{k0}=0,
\end{equation}
because the closed ground state is even.
Hence,
\begin{equation}
    \lim_{T_i\to0^+}a_k(T_i)=0
    \qquad
    (k:\ {\rm odd}).
    \label{eq:ak_lowT_odd}
\end{equation}
For an even resonance mode with $k\neq0$, the low-temperature limit is not identically zero.  For a weak CAP whose support is sufficiently separated from the localized core states, however, $|O_{k0}|^2$ can be very small.

To see the origin of this small overlap, ordinary non-Hermitian perturbation theory gives, for an isolated $k$-th excited state,
\begin{equation}
    |\phi_k\rangle
    =
    |k\rangle
    -i\sum_{\ell\neq k}
    \frac{
        \langle\ell|\hat W|k\rangle
    }{
        E_k-E_\ell
    }
    |\ell\rangle
    +
    \mathcal O(\|W\|^2).
    \label{eq:phik_perturbation}
\end{equation}
Consequently, for an even mode the projection onto the closed ground state is controlled at leading order by the boundary matrix element
\begin{equation}
    \frac{
        \langle0|\hat W|k\rangle
    }{
        E_k-E_0
    },
\end{equation}
and is suppressed when both closed states have exponentially small amplitudes in the absorbing region.

\subsubsection{High-temperature limit}
For the continuum problem on a finite interval, the closed Hamiltonian has an infinite discrete spectrum and
\begin{equation}
    Z(T_i)\longrightarrow\infty
    \qquad
    (T_i\to\infty).
\end{equation}
On the other hand, completeness of the orthonormal eigenstates of $\hat H_0$ gives
\begin{equation}
    \sum_{m=0}^{\infty}|O_{km}|^2
    =
    \sum_{m=0}^{\infty}
    |\langle\chi_k|m\rangle|^2
    =
    \langle\chi_k|\chi_k\rangle
    <\infty.
    \label{eq:parseval_ak}
\end{equation}
The same-parity part of this sum is therefore also finite.
Hence,
\begin{equation}
    \lim_{T_i\to\infty}a_k(T_i)=0.
    \label{eq:ak_highT}
\end{equation}

At a fixed finite grid size, the numerical Hilbert space is finite-dimensional and the limiting value is instead
\begin{equation}
    \frac{
        \langle\chi_k|\chi_k\rangle
    }{
        N_{\mathrm{grid}}
    }.
\end{equation}
Therefore, the continuum high-temperature limit should be checked by increasing the grid size before taking $T_i$ arbitrarily large.

\subsubsection{Intermediate-temperature activation}
At an intermediate temperature, the closed state $|k\rangle$ acquires a finite Gibbs weight.
For any $T^*>0$, positivity of every term in Eq.~\eqref{Eq18} gives the lower bound
\begin{equation}
    a_k(T^*)
    \geq
    P_k(T^*)|O_{kk}|^2,
    \qquad
    P_k(T^*)
    =
    \frac{
        e^{-E_k/(k_BT^*)}
    }{
        Z(T^*)
    }
    >0.
    \label{eq:ak_intermediate_bound}
\end{equation}

In the weak-CAP regime, Eq.~\eqref{eq:phik_perturbation} implies that the open resonance mode $|\phi_k\rangle$ is continuously connected to the closed state $|k\rangle$, so that $|O_{kk}|$ remains of order unity.
For an even mode with $k\neq0$, $|O_{k0}|$ can simultaneously be small.  
A simple sufficient condition for the intermediate value to exceed the low-temperature limit is therefore
\begin{equation}
    P_k(T^*)|O_{kk}|^2
    >
    |O_{k0}|^2.
    \label{eq:sufficient_peak_condition}
\end{equation}
If Eq.~\eqref{eq:sufficient_peak_condition} holds, then
\begin{equation}
    a_k(T^*)
    >
    \lim_{T_i\to0^+}a_k(T_i).
\end{equation}

For an odd resonance mode, $O_{k0}=0$ by parity, so that Eq.~\eqref{eq:sufficient_peak_condition} is automatically satisfied whenever
\begin{equation}
    P_k(T^*)|O_{kk}|^2>0.
\end{equation}
Thus, provided that the $k$-th closed state couples to the corresponding resonance mode, $a_k(T_i)$ is activated from zero as the temperature increases.

The sufficient condition~\eqref{eq:sufficient_peak_condition} is valid independently of the nodal structure of the resonance eigenfunction and can be tested directly from the calculated overlaps.

\subsubsection{Existence of a finite-temperature maximum}
The function $a_k(T_i)$ is continuous for $T_i>0$ and has the endpoint limits given in Eqs.~\eqref{eq:ak_lowT} and \eqref{eq:ak_highT}.
Suppose that there exists a temperature $T^*>0$ satisfying Eq.~\eqref{eq:sufficient_peak_condition}.
Then $a_k(T^*)$ exceeds its low-temperature limit, while $a_k(T_i)$ approaches zero at high temperature.
It follows that $a_k(T_i)$ attains at least one maximum at a finite temperature
\begin{equation}
    T_{\mathrm{peak},k}\in(0,\infty).
\end{equation}
At an isolated differentiable maximum,
\begin{equation}
    \left.
    \frac{da_k}{dT_i}
    \right|_{T_i=T_{\mathrm{peak},k}}
    =0,
    \qquad
    \left.
    \frac{d^2a_k}{dT_i^2}
    \right|_{T_i=T_{\mathrm{peak},k}}
    \leq0.
    \label{eq:ak_peak_derivatives}
\end{equation}
For a nondegenerate maximum, the second inequality is strict.

We have therefore established the following parameter-dependent result:

\begin{quote}
For a resonance mode $k\neq0$, if there exists an intermediate temperature $T^*$ such that
\begin{equation*}
    P_k(T^*)|O_{kk}|^2>|O_{k0}|^2,
\end{equation*}
then the diagonal spectral coefficient $a_k(T_i):=a_{k,k}(T_i)$ necessarily exhibits at least one finite-temperature maximum.
For odd $k$, the right-hand side vanishes identically by parity.
\end{quote}

Thus, the non-monotonic temperature dependence is not restricted to the second even mode.
The same mechanism can occur for any nontrivial resonance channel: thermal excitation increases the weight of a closed state strongly coupled to the corresponding resonance, while the normalization by the partition function suppresses $a_k(T_i)$ again in the high-temperature limit.

The numerical results presented below show which resonance channels exhibit such non-monotonic activation for the parameters studied in this work.
These temperature-dependent spectral weights, together with the different decay rates $\Gamma_k$, determine the relaxation behavior of
\begin{equation}
    S(t,T_i)
    =
    \sum_k
    a_k(T_i)e^{-\Gamma_k t/\hbar}.
\end{equation}
Their competition provides the spectral mechanism underlying the Mpemba crossing discussed below.

\subsubsection{Dependence on the system size and parameter regime}
For the parameter set studied here, the numerically determined positions of the peaks in $a_k(T_i)$ for the relevant low-lying resonance modes become insensitive, within numerical resolution, to further increases of $L$ once the absorbing boundaries are sufficiently separated from the core region.
This behavior is consistent with the localization of the relevant closed states and with the weak dependence of the overlaps $O_{km}$ on remote boundaries.
It should be understood as asymptotic boundary insensitivity in the investigated regime, rather than exact independence for arbitrary $L$.

The non-monotonic peaks can weaken or disappear outside this regime.
In particular, an excessively strong or abrupt CAP can produce appreciable reflection and strongly deform the resonance modes; a shallow double well can remove the separation between localized core states and rapidly leaking excited states; and a boundary placed too close to the wells can increase the low-temperature overlaps $|O_{k0}|$ for even modes and invalidate the sufficient condition~\eqref{eq:sufficient_peak_condition}.
These observations explain why the peaks can be robust over a broad useful parameter range but are not guaranteed for every choice of the potential and absorbing layer.

\subsection{Trace Distance of the Conditional State}
\label{sec:trace_distance}
The spectral relaxation measure introduced above characterizes the decay-weighted distribution of the diagonal resonance weights.
To examine whether the same anomalous ordering also appears in the relaxation of the full quantum state, we consider the normalized conditional density operator
\begin{equation}
    \hat{\rho}_{\rm cond}(t,T_i)
    :=
    \frac{\hat{\rho}_c(t,T_i)}
    {\mathrm{Tr}[\hat{\rho}_c(t,T_i)]}.
    \label{eq:rho_cond}
\end{equation}
Here, $\hat{\rho}_c(t,T_i)$ is the subnormalized no-jump density operator introduced in Eq.~\eqref{rho_c(t)}.  
The normalization in Eq.~\eqref{eq:rho_cond} removes the overall loss of probability and therefore describes the quantum state of the particle conditioned on its survival up to time $t$.

We next define the asymptotic conditional state.  
Suppose that the resonance mode $k=0$ has the smallest decay rate and is nondegenerate,
\begin{equation}
    \Gamma_0 < \Gamma_k
    \qquad (k\neq0).
    \label{eq:unique_slowest}
\end{equation}
From the full spectral expansion in Eq.~\eqref{eq:rho_full_spectral_main}, the long-time behavior of the subnormalized density operator is then dominated by the slowest resonance,
\begin{equation}
    \hat{\rho}_c(t,T_i)
    \simeq
    a_{00}(T_i)
    e^{-\Gamma_0t/\hbar}
    |\phi_0\rangle\langle\phi_0|.
    \label{eq:rho_longtime}
\end{equation}
Consequently, after normalization,
\begin{equation}
    \hat{\rho}_{\rm cond}(t,T_i)
    \longrightarrow
    \hat{\rho}_{\rm QS}
    :=
    \frac{|\phi_0\rangle\langle\phi_0|}
    {\langle\phi_0|\phi_0\rangle}
    \qquad
    (t\to\infty).
    \label{eq:rho_QS}
\end{equation}
The state $\hat{\rho}_{\rm QS}$ is the asymptotic quasi-stationary state of the surviving ensemble.  
Importantly, under the assumption in Eq.~\eqref{eq:unique_slowest}, it is independent of the initial preparation temperature.

We quantify the approach to this asymptotic state by the trace distance
\begin{equation}
    D_{\rm QS}(t,T_i)
    :=
    \frac{1}{2}
    \left\|
        \hat{\rho}_{\rm cond}(t,T_i)
        -
        \hat{\rho}_{\rm QS}
    \right\|_1,
    \label{eq:DQS}
\end{equation}
where
\begin{equation}
    \|A\|_1
    :=
    \mathrm{Tr}\sqrt{A^\dagger A}
\end{equation}
denotes the trace norm.  
Since both $\hat{\rho}_{\rm cond}$ and $\hat{\rho}_{\rm QS}$ are normalized
density operators,
\begin{equation}
    0\leq D_{\rm QS}(t,T_i)\leq1,
\end{equation}
and Eq.~\eqref{eq:rho_QS} implies
\begin{equation}
    \lim_{t\to\infty}D_{\rm QS}(t,T_i)=0.
    \label{eq:DQS_longtime}
\end{equation}

At the initial time, $\mathrm{Tr}\hat{\rho}_i(T_i)=1$, and hence
\begin{equation}
    D_{\rm QS}(0,T_i)
    =
    \frac{1}{2}
    \left\|
        \hat{\rho}_i(T_i)-\hat{\rho}_{\rm QS}
    \right\|_1.
    \label{eq:DQS_initial}
\end{equation}
Unlike the distance from the vacuum state in the enlarged particle--vacuum Hilbert space, this initial distance generally depends on the preparation temperature $T_i$.  
It therefore allows us to compare states that are initially located at different distances from the asymptotic conditional state.

For two initial temperatures $T_h>T_c$, we define
\begin{equation}
    \Delta D_{\rm QS}(t)
    :=
    D_{\rm QS}(t,T_h)-D_{\rm QS}(t,T_c).
    \label{eq:DeltaDQS}
\end{equation}
If
\begin{equation}
    \Delta D_{\rm QS}(0)>0
\end{equation}
but $\Delta D_{\rm QS}(t)$ changes sign at a finite time, the state that is initially farther from the asymptotic state becomes closer at a later time.  
Such an inversion would constitute a trace-distance Mpemba crossing of the conditional quantum state.

In contrast to the diagonal spectral measure $S(t,T_i)$, the trace distance in Eq.~\eqref{eq:DQS} is evaluated from the full conditional density operator and therefore contains the effects of both diagonal and off-diagonal coefficients $a_{k,j}(T_i)$.  
Thus, $D_{\rm QS}$ tests whether the spectral Mpemba crossing identified from $S$ is accompanied by an anomalous acceleration of the complete normalized quantum-state relaxation.

We finally note that the usual contractivity of the trace distance under a common completely positive trace-preserving map does not directly imply monotonicity of $D_{\rm QS}(t,T_i)$, because the conditioning and normalization in Eq.~\eqref{eq:rho_cond} make the effective evolution nonlinear.  
The behavior of $D_{\rm QS}$ must therefore be examined explicitly, as we do numerically below.

\section{Numerical confirmation}\label{Sec:confirmation}
\subsection{Setup}
\begin{figure}[hbtp]
    \centering
    \includegraphics[width=0.75\linewidth]{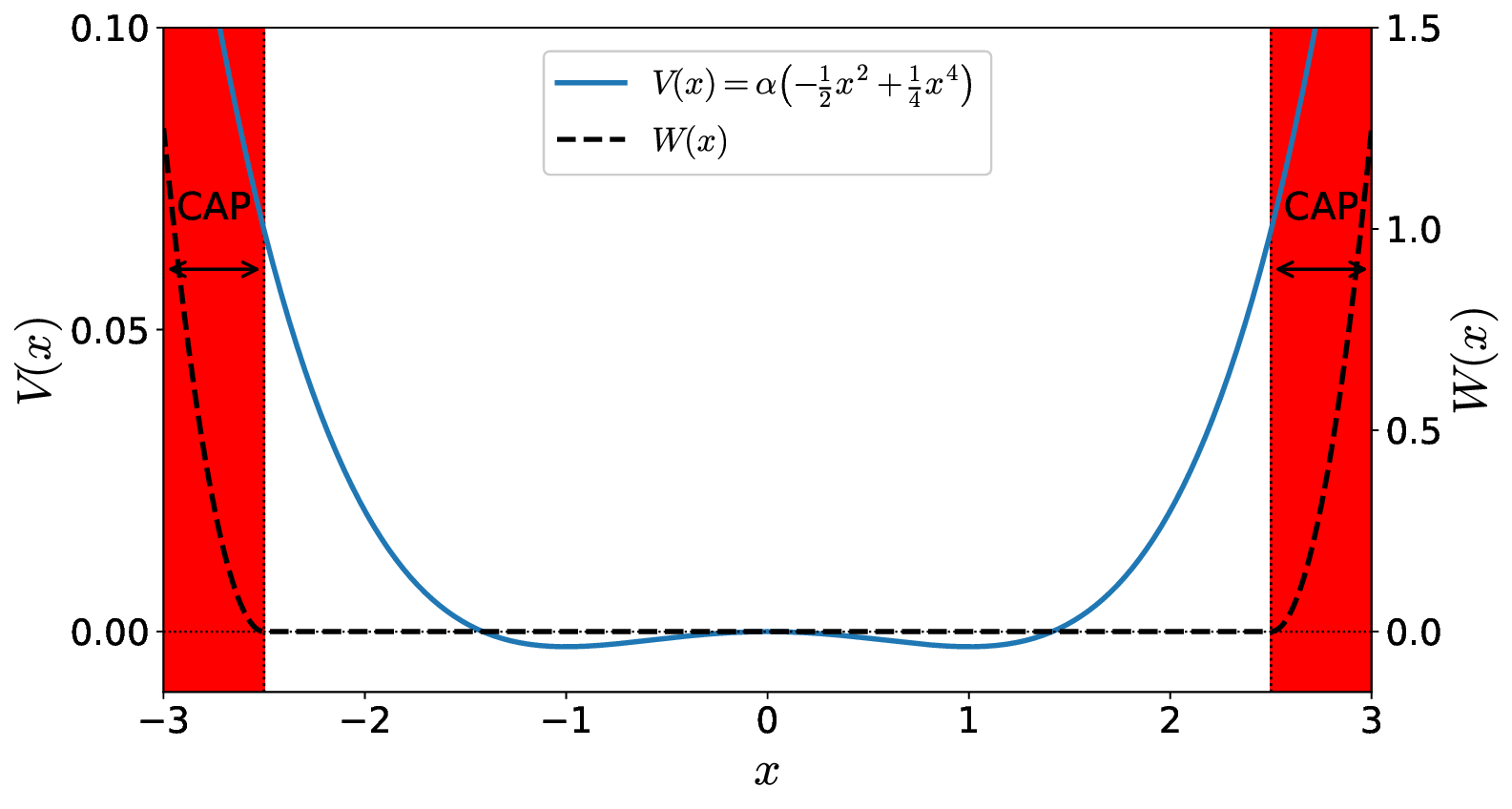}
    \caption{Schematic illustration of the potential $V(x)=\alpha(-x^2/2+x^4/4)$ and the complex absorbing potential (CAP) $W(x)$ given by Eq.~\eqref{W(x)} for $L=3$, $x_{\mathrm{abs}}=0.5$, $\eta=5$, $\alpha=0.01$, and the CAP exponent $n=2$.
    The shaded regions indicate the CAP layers of width $x_{\mathrm{abs}}$ near the boundaries.
    }
    \label{fig:potential}
\end{figure}
In this section, we demonstrate the validity of the theoretical argument in the previous section using the symmetric potential
\begin{equation}
    V(x)
    =\alpha\left(-\frac{1}{2}x^2+\frac{1}{4}x^4\right),
\end{equation}
where $\alpha>0$ sets the overall energy scale of the double-well potential (see Fig.~\ref{fig:potential}).
The locations of the two minima, $x=\pm1$, which correspond to the unit length of this system, and the barrier position, $x=0$, are independent of $\alpha$, whereas the barrier height relative to the minima is given by $\Delta V=\alpha/4$.

We discretize the spatial coordinate using a uniform finite-difference grid and numerically diagonalize both the Hermitian Hamiltonian $H_0$ and the non-Hermitian Hamiltonian $H_{\mathrm{eff}}$.
See also Appendix~\ref{sec:numerical} for the detailed explanation.
To compute the resonance spectrum, we discretize the coordinate representation of the effective Hamiltonian on a uniform spatial grid consisting of $N=1200$ points in the interval $[-L,L]$.
The eigenstates of the closed system are first obtained by solving Eq.~\eqref{eigen_H_0}.
The coordinate representation of $\hat{H}_0$ is discretized on the uniform spatial grid, and the resulting matrix eigenvalue problem is solved numerically using a finite-difference approximation of the kinetic-energy operator.

Subsequently, the resonance states are obtained from Eq.~\eqref{eigenequation}.
In the numerical implementation, the right eigenvectors correspond to the coordinate-space wavefunctions
\begin{equation}
    \phi_k(x)
    := \langle x|\phi_k\rangle,
\end{equation}
evaluated at the discrete grid points.
The corresponding left eigenstates are expressed as
\begin{equation}
    \chi_k(x)
    := \langle x|\chi_k\rangle.
\end{equation}
The overlap matrix elements are evaluated as
\begin{equation}
    O_{kn}
    = \langle \chi_k|n\rangle,
\end{equation}
from which the spectral coefficients are constructed.
In particular, the generalized coefficients $a_{k,j}(T_i)$ are evaluated from Eq.~\eqref{eq:akj_main} using the resulting biorthogonal eigenbasis, and their diagonal elements $a_k(T_i)=a_{k,k}(T_i)$ give the spectral weights entering the spectral relaxation measure.

Unless otherwise stated, we fix the parameters as
\begin{equation}\label{parameters}
    L=3,\quad
    \eta=5,\quad
    x_{\mathrm{abs}}=0.5,\quad
    \alpha=0.01,\quad
    n=2.
\end{equation}

\subsection{Spectral coefficients}
\begin{figure}[hbtp]
    \centering
    \includegraphics[width=\linewidth]{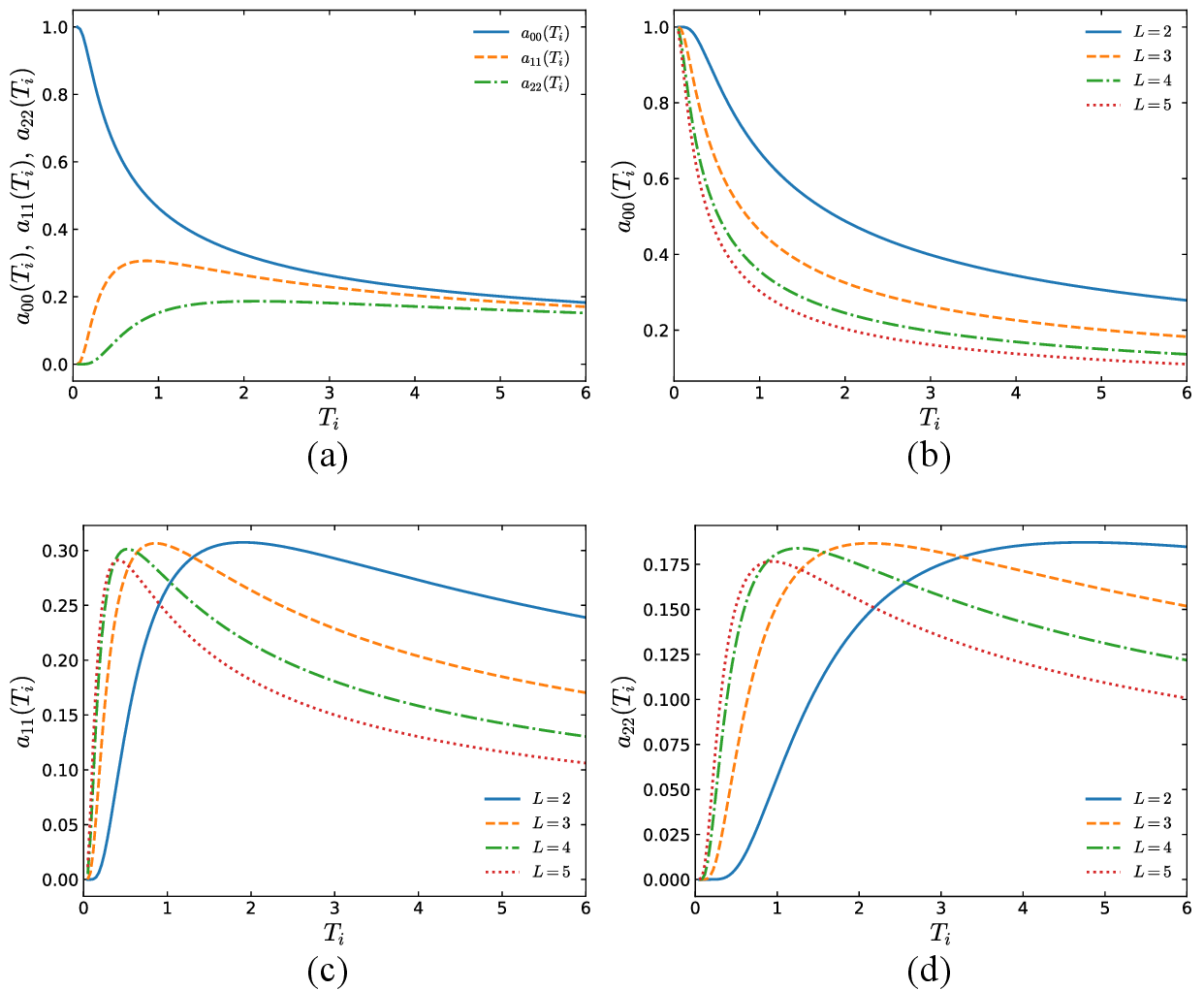}
    \caption{
    (a) Initial-temperature dependence of $a_{00}(T_i)$, $a_{11}(T_i)$, and $a_{22}(T_i)$ for $L=3$.
    (b--d) System-size dependence of (b) $a_{00}(T_i)$, (c) $a_{11}(T_i)$, and (d) $a_{22}(T_i)$ for $L=2$ (solid lines), $3$ (dashed lines), $4$ (dot-dashed lines), and $5$ (dotted lines).
    The remaining parameters are $\eta=5$, $x_\mathrm{abs}=0.5$, and $\alpha=0.01$.
    }
    \label{fig:an_vs_T}
\end{figure}

Figure~\ref{fig:an_vs_T}(a) shows the initial-temperature dependence of the diagonal spectral coefficients $a_k(T_i)=a_{k,k}(T_i)$ for $k=0$, $1$, and $2$.
For this parameter set, the excited-channel weights $a_1(T_i)$ and $a_2(T_i)$ exhibit a maximum at an intermediate temperature, whereas $a_0(T_i)$ decreases monotonically with increasing $T_i$ over the displayed range.
This is a numerical property of the chosen model and is not asserted to be universal.

Figure~\ref{fig:an_vs_T}(b--d) shows $a_{00}(T_i)$, $a_{11}(T_i)$, and $a_{22}(T_i)$ for several system sizes.
We can observe that $a_{00}(T_i)$ is a monotonically decreasing function of $T_i$.
In addition, within the displayed numerical resolution, the peak positions of $a_{11}(T_i)$ and $a_{22}(T_i)$ shift toward lower temperatures as the system size $L$ increases.

\subsection{Spectral Relaxation Measure}
\begin{figure}[hbtp]
    \centering
    \includegraphics[width=0.49\linewidth]{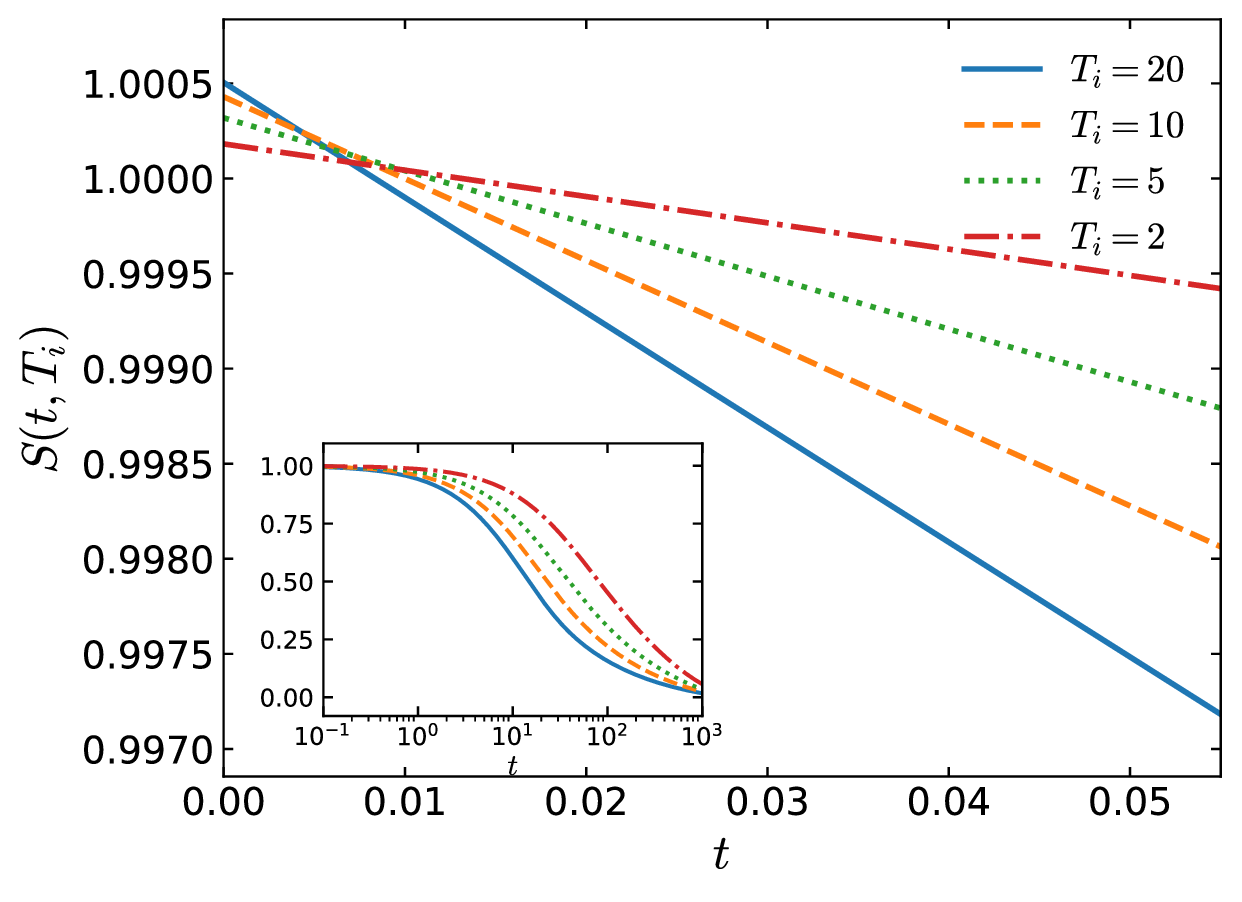}
    \caption{
    Time evolution of the spectral relaxation measure $S(t,T_i)$ for $T_i=2$ (dot-dashed line), $5$ (dotted line), $10$ (dashed line), and $20$ (solid line) for $L=3$, $\eta=5$, $x_\mathrm{abs}=0.5$, and $\alpha=0.01$.
    The main panel highlights the short-time regime in which the relaxation curves cross, while the inset shows the corresponding long-time relaxation over $10^{-1}\le t\le10^3$ on a logarithmic
    time scale.
    }
    \label{fig:survival}
\end{figure}

Figure~\ref{fig:survival} shows the time evolution of the spectral relaxation measure $S(t,T_i)$ for four different initial temperatures, $T_i=2$, $5$, $10$, and $20$.
Although each curve decreases monotonically with time, the curves exhibit crossings within a finite-time interval, indicating a reversal of their relative ordering.
This behavior provides a direct numerical demonstration of the spectral Mpemba crossing defined in Sec.~\ref{sec:spectral_crossing}.

\begin{figure}[hbtp]
    \centering
    \includegraphics[width=\linewidth]{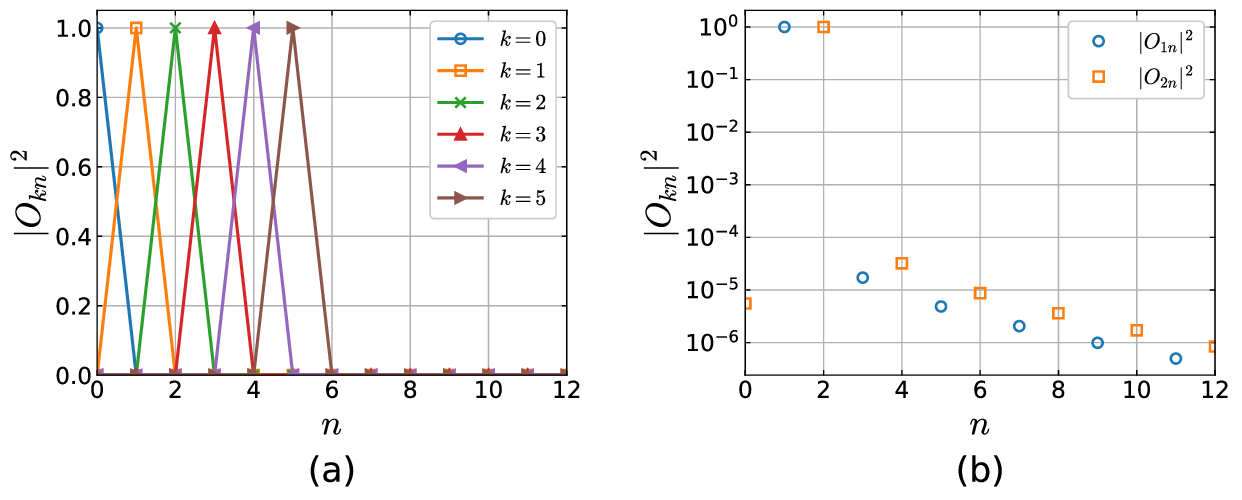}
    \caption{
    (a) Plot of $|O_{kn}|^2$ against $n$ for various $k$ for $L=3$, $\eta=5$, $x_\mathrm{abs}=0.5$, and $\alpha=0.01$.
    (b) Semi-logarithmic plots of the same-parity overlaps $|O_{1n}|^2$ (open circles) and $|O_{2n}|^2$ (open squares) against $n$ for the same set of parameters.
    }
    \label{fig:overlap}
\end{figure}

Figure~\ref{fig:overlap}(a) plots the profiles of the overlap matrix $|O_{kn}|^2$ for various $k$ and $n$.
The overlap matrix is strongly concentrated near the diagonal, indicating that each low-lying resonance mode has its largest overlap with the corresponding closed-system eigenstate.
Smaller off-diagonal components are also present, and their relative contribution becomes more visible for higher modes.

Figure~\ref{fig:overlap}(b) shows the same-parity overlaps $|O_{1n}|^2$ and $|O_{2n}|^2$ on a semi-logarithmic scale.
Because the $k=1$ and $k=2$ resonance modes have odd and even parity, respectively, $O_{1n}$ is nonzero only for odd $n$, whereas $O_{2n}$ is nonzero only for even $n$.
In both cases, the overlap is largest near the corresponding diagonal element and rapidly decreases as $n$ moves away from it.
This behavior provides a numerical illustration of the parity selection rule and the strong correspondence between low-lying closed-system eigenstates and resonance modes.

\begin{figure}[hbtp]
    \centering
    \includegraphics[width=0.49\linewidth]{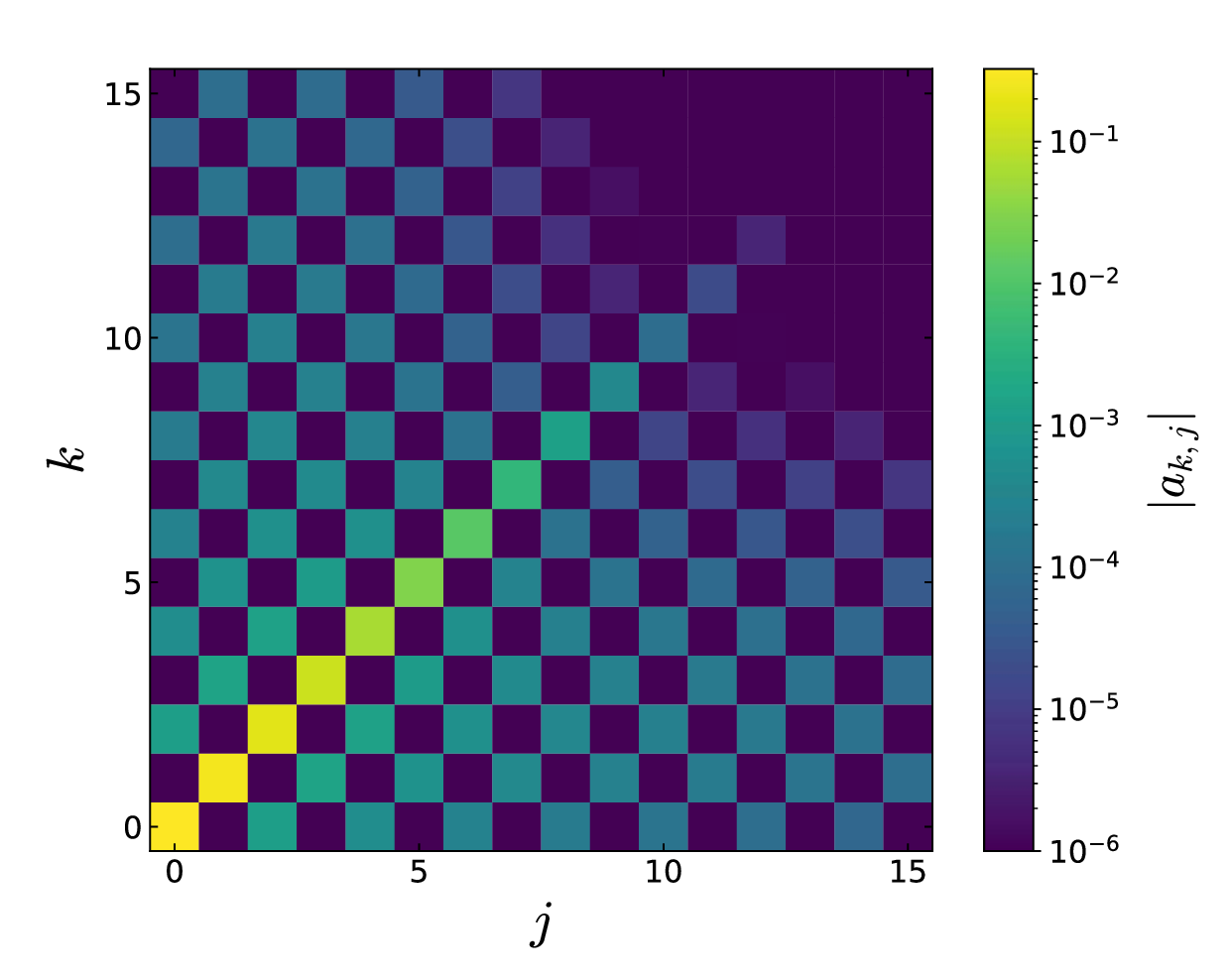}
    \caption{
    Heatmap of $|a_{k,j}(T_i)|$ for $T_i=2$, $L=3$, $\eta=5$, $x_\mathrm{abs}=0.5$, and $\alpha=0.01$.
    }
    \label{fig:A}
\end{figure}

Figure~\ref{fig:A} shows the heatmap of $|a_{k,j}(T_i)|$ defined in Eq.~\eqref{eq:akj_main} for $T_i=2$ using the parameters in Eq.~\eqref{parameters}.
The diagonal elements are dominant, while finite off-diagonal elements are also present between modes belonging to the same parity sector.
A characteristic checkerboard pattern is clearly visible.
In particular, the elements with odd $|k-j|$ vanish because the two resonance modes then have opposite parity.
This numerical structure is consistent with the parity selection rule in Eq.~\eqref{eq:akj_parity}.

\subsection{Trace-Distance Relaxation}
\label{sec:numerical_trace_distance}
We next examine whether the spectral Mpemba crossing observed in $S(t,T_i)$ is accompanied by an anomalous relaxation of the complete normalized conditional quantum state.
For this purpose, we evaluate the trace distance $D_{\rm QS}(t,T_i)$ defined in Eq.~\eqref{eq:DQS}.
In contrast to $S(t,T_i)$, which is constructed only from the diagonal resonance weights $a_{k,k}(T_i)$, $D_{\rm QS}(t,T_i)$ is calculated from the full conditional density matrix and therefore retains the influence of both the diagonal and off-diagonal coefficients $a_{k,j}(T_i)$.

\begin{figure}[hbtp]
    \centering
    \includegraphics[width=0.49\linewidth]{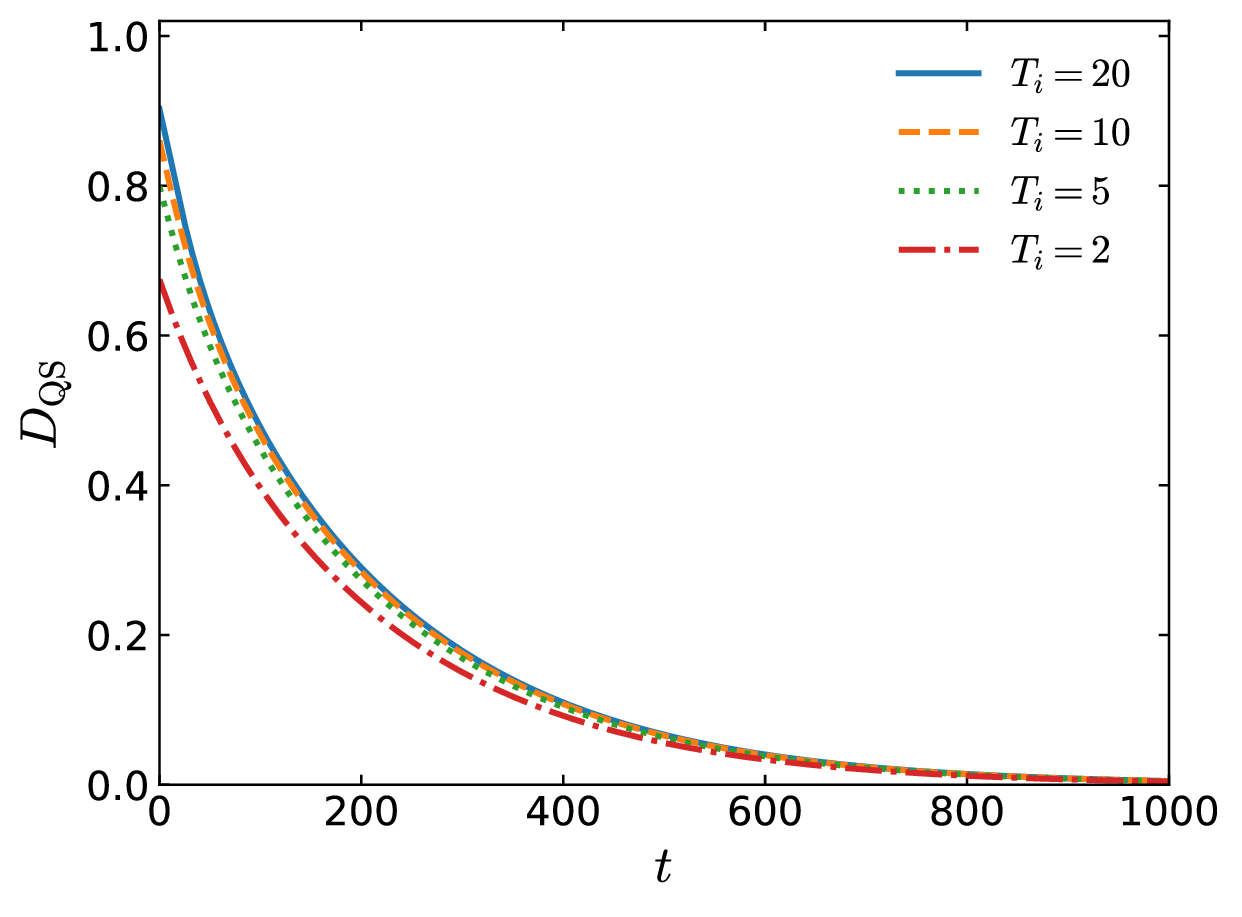}
    \caption{
    Time evolution of the trace distance $D_{\rm QS}(t,T_i)$ between the normalized conditional state $\hat{\rho}_{\rm cond}(t,T_i)$ and its asymptotic quasi-stationary state $\hat{\rho}_{\rm QS}$ for $T_i=2$ (dot-dashed line), $5$ (dotted line), $10$ (dashed line), and $20$ (solid line).
    The parameters are fixed at $L=3$, $\eta=5$, $x_\mathrm{abs}=0.5$, and $\alpha=0.01$.
    }
    \label{fig:D_QS}
\end{figure}

Figure~\ref{fig:D_QS} shows $D_{\rm QS}(t,T_i)$ for $T_i=2$, $5$, $10$, and $20$.
At $t=0$, the trace distance clearly depends on $T_i$: the higher-temperature preparations lie farther from the asymptotic conditional state $\hat{\rho}_{\rm QS}$.
This behavior is natural because increasing $T_i$ populates excited closed-system states more strongly, producing an initial conditional state with a larger contribution from resonance channels distinct from the longest-lived mode.

For all temperatures shown in Fig.~\ref{fig:D_QS}, $D_{\rm QS}(t,T_i)$ decreases toward zero as time increases, consistent with the convergence $\hat{\rho}_{\rm cond}(t,T_i)\to\hat{\rho}_{\rm QS}$.
This figure also confirms the overall decay toward the quasi-stationary state over the full displayed time interval.

Most importantly, the ordering of the curves is preserved: a state that starts farther from $\hat{\rho}_{\rm QS}$ remains farther throughout the displayed relaxation process.
In particular, the curves remain ordered from top to bottom as $T_i=2$, $5$, $10$, and $20$.
We therefore find no sign change of
\begin{equation}
    \Delta D_{\rm QS}(t)
    =
    D_{\rm QS}(t,T_h)-D_{\rm QS}(t,T_c)
\end{equation}
for the temperature pairs examined here.

Thus, no trace-distance Mpemba crossing is observed for the present parameter set.
This result is in clear contrast to the crossing of the diagonal spectral relaxation measure $S(t,T_i)$ shown in Fig.~\ref{fig:survival}.
The difference demonstrates that the anomalous ordering of $S(t,T_i)$ does not imply a universal acceleration of the full conditional quantum-state relaxation.
Rather, the effect found here is specifically associated with the temperature-dependent redistribution of resonance weights followed by their mode-dependent decay filtering.
In this sense, the present result supports the interpretation of the observed phenomenon as a \emph{spectral Mpemba effect}.

\subsection{Comparison between Spectral and Trace-Distance Relaxation}
\label{sec:comparison_relaxation}
The numerical results demonstrate that the spectral relaxation measure $S(t,T_i)$ and the trace distance $D_{\rm QS}(t,T_i)$ probe distinct aspects of the non-Hermitian relaxation.
The former exhibits a finite-time reversal of the ordering between different thermal preparations, whereas the latter preserves their ordering for the parameter regime studied here.
The observed anomalous relaxation should therefore be understood as a \emph{spectral Mpemba effect}, rather than as a universal acceleration of the complete conditional quantum-state dynamics.

\section{Discussion}\label{Sec:Discussion}
In this section, we examine the physical mechanisms underlying the observed spectral Mpemba effect, assess the role of symmetry and system size, and contextualize our real-space non-Hermitian framework within the broader landscape of open quantum thermodynamics.

\subsection{Physical Mechanism: Gibbs Weights vs. Non-Hermitian Overlaps}
The continuous, coordinate-space non-Hermitian formulation developed in this work offers a concrete physical perspective on anomalous quantum relaxation. 
While quantum Mpemba effects in discrete spin networks or abstract state spaces are typically characterized via state-space metrics or entanglement measures, our continuous framework directly links real-space wavefunction profiles and localized boundary loss to the spectral relaxation dynamics.

The complex absorbing potential $W(x)$ creates localized dissipation channels near the terminal boundaries $x = \pm L$. Higher-energy eigenstates of the closed Hamiltonian $\hat{H}_0$ possess greater spatial extent and kinetic energy, allowing them to penetrate deeper into the absorbing layers than deeply bound, core-localized ground states. 
The initial preparation temperature $T_i$ controls the Gibbs population of these closed states $P_n(T_i)$, which project onto the non-Hermitian resonance modes $|\phi_k\rangle$ via the biorthogonal overlaps $O_{kn} = \langle \chi_k | n \rangle$. 

As shown in Fig.~\ref{fig:overlap}, the overlap matrix elements $|O_{kn}|^2$ are strongly concentrated along the diagonal ($k \approx n$), reflecting a robust one-to-one correspondence between closed states and open resonance modes, while off-diagonal overlaps decay rapidly with increasing $|k-n|$.
Crucially, the non-monotonic temperature dependence observed in the diagonal spectral weights $a_k(T_i)$, specifically $a_1(T_i)$ and $a_2(T_i)$, is a direct manifestation of the competition between thermal population and spatial mode overlap:
\begin{enumerate}
    \item At low temperatures ($T_i\to0^+$), only the closed-system ground state is populated, and hence $a_k(T_i)\to|O_{k0}|^2$.
    For odd-parity resonance modes, this limit vanishes exactly by reflection symmetry. For even-parity excited modes, it is generally finite but can be strongly suppressed when the absorbing region is well separated from the localized ground state.
    \item As $T_i$ increases, thermal activation populates the corresponding excited closed state $|k\rangle$, causing $a_k(T_i)$ to grow rapidly due to the strong diagonal coupling $O_{kk} \approx 1$.
    \item At high temperatures ($T_i \to \infty$), the growth of the partition function $Z(T_i)$ suppresses individual mode occupations, driving $a_k(T_i) \to 0$.
\end{enumerate}
For appropriate resonance channels and parameter regimes, this trade-off can generate a finite-temperature maximum $T_{\rm peak,k}$ in $a_k(T_i)$, providing the energy-dependent redistribution of initial weight required for a spectral Mpemba
crossing., providing the necessary energy-dependent redistribution of initial weight across distinct decay channels $\Gamma_k$.

\subsection{Spectral Relaxation vs. State-Space Trajectory}
A fundamental result of this work is the sharp distinction between the diagonal spectral relaxation measure $S(t,T_i)$ and the full state-space evolution tracked by the conditional trace distance $D_{\text{QS}}(t,T_i)$.

The measure $S(t,T_i)$ intentionally focuses on the diagonal resonance decay factor $\sum_k a_k(T_i) e^{-\Gamma_k t / \hbar}$. 
Although $S(t,T_i)$ decays monotonically for any individual initial temperature, curves corresponding to different initial temperatures $T_h > T_c$ display finite-time crossings $S(t_\times, T_h) = S(t_\times, T_c)$. 
This behavior arises because the initial thermal state prepared at $T_h$ places optimal spectral weight into rapidly decaying resonance modes, allowing its total diagonal spectral content to drop below that of an initially colder preparation $T_c$.

Conversely, the trace distance $D_{\text{QS}}(t,T_i)$ evaluates the full normalized conditional state $\hat{\rho}_{\text{cond}}(t) = \hat{\rho}_c(t) / \operatorname{Tr}[\hat{\rho}_c(t)]$ relative to the asymptotic quasi-stationary state $\hat{\rho}_{\text{QS}}$, incorporating all same-parity off-diagonal interference terms $a_{k,j}(T_i) e^{-i(\lambda_k - \lambda_j^*)t/\hbar}$. For the parameter regimes investigated in Sec.~\ref{Sec:confirmation}, $D_{\text{QS}}(t,T_i)$ decreases monotonically toward zero for all initial temperatures without experiencing any crossing.

This absence of crossing in $D_{\text{QS}}(t,T_i)$ prevents a common misinterpretation: the quantum Mpemba phenomenon observed here is \emph{not} a universal state-space acceleration of the entire density operator. 
Instead, it is strictly a \emph{spectral Mpemba effect}, wherein anomalous ordering is restricted to specific observable subspaces or spectral relaxation measures. 
This result highlights that in open quantum systems, whether an anomalous relaxation effect is observed depends crucially on the physical observable chosen to probe the system.

We also distinguish the present finite-time spectral crossing from the strong Mpemba effect, in which the coefficient of the slowest relevant relaxation mode vanishes and the approach to stationarity is exponentially accelerated~\cite{Carollo21}.
In the present formulation,
\begin{equation}
    a_k(T_i)
    =
    \sum_n P_n(T_i)
    |\langle\chi_k|n\rangle|^2
\end{equation}
is a sum of nonnegative contributions.
At any finite temperature, an exact zero of $a_k(T_i)$ would therefore require all overlaps with thermally occupied states in the relevant symmetry sector to vanish.
Such a cancellation is not generic in the present continuous CAP model and is not observed for the parameter regime considered here.
The effect reported in this work is instead a finite-time ordering reversal caused by a temperature-dependent redistribution of positive spectral weights among resonance modes with different decay rates.

Beyond the canonical thermal preparations considered here, the same spectral construction can also be extended to other Mpemba-like protocols.
For an initial state $\hat\rho_i(\xi)$ parameterized by a general preparation variable $\xi$, one may define
\begin{equation}
    a_{k,j}(\xi)
    =
    \langle\chi_k|\hat\rho_i(\xi)|\chi_j\rangle,
    \qquad
    a_k(\xi)=a_{k,k}(\xi).
\end{equation}
The subsequent filtering by the resonance decay factors remains unchanged.
This provides a possible route for applying the present spectral framework to nonthermal or time-delayed preparation protocols, such as the recently proposed Descartes protocol~\cite{Santos2026}.
A systematic investigation of such protocols is left for future work.

\subsection{Role of Reflection Symmetry and Boundary Scaling}

The exact parity selection rule $a_{k,j}(T_i) = 0$ for opposite-parity modes relies on the strict reflection symmetry of both the potential $V(-x) = V(x)$ and the absorbing profile $W(-x) = W(x)$. 
If reflection symmetry is broken (e.g., in asymmetric double wells or tilted potentials), the parity selection rule breaks down, allowing cross-parity off-diagonal elements to become non-zero. 
Nevertheless, the general mechanism—thermal redistributing of initial weights over non-Hermitian modes with distinct decay rates—does not rely on exact spatial reflection symmetry. 
Provided that $a_k(T_i)$ exhibits non-monotonic thermal activation in an asymmetric potential, spectral Mpemba crossings can still occur.

Furthermore, we contrast our quantum results with classical boundary-driven Mpemba effects observed in overdamped Langevin dynamics~\cite{Yue26, Yue_long26}. 
In classical systems, the outer boundary position $L$ directly dictates the escape rates and crossing times. 
In our quantum formulation, as the computational boundary $L$ increases beyond the spatial extent of the core potential $V(x)$, the peak positions of the selected $a_k(T_i)$ shown in Fig.~\ref{fig:an_vs_T} become only weakly dependent on $L$. 
This boundary insensitivity reflects the strong spatial localization of the low-lying closed states within the core wells: once $L$ is sufficiently large, remote boundary sinks no longer significantly alter the core overlaps $O_{kn}$.

\section{Conclusion}\label{Sec:Conclusion}

In this paper, we have formulated and analyzed a continuous coordinate-space realization of the quantum tunneling Mpemba effect in an open double-well system subject to localized boundary dissipation. 
By embedding complex absorbing potentials into the continuous spatial domain and applying a biorthogonal spectral decomposition, we established a rigorous mapping from microscopic boundary loss to non-Hermitian mode dynamics.

Our main findings are summarized as follows:
1) \textit{Biorthogonal Mode Structure and Parity Selection:} 
We derived the full spectral expansion of the subnormalized conditional density matrix $\hat{\rho}_c(t)$. 
For spatial reflection-symmetric potentials and loss profiles, we proved that off-diagonal spectral elements $a_{k,j}(T_i)$ vanish identically between opposite-parity modes, while same-parity cross-mode terms remain finite, yielding a characteristic checkerboard structure.
2) \textit{Non-Monotonic Spectral Activation:} 
We derived a parameter-dependent sufficient condition for a diagonal spectral coefficient $a_k(T_i)$ to exhibit a finite-temperature maximum. For the parameter set studied numerically, selected excited spectral coefficients display such non-monotonic temperature dependence.
This non-monotonicity stems from the competition between the Gibbs Boltzmann factor $P_n(T_i)$ and the spatial overlap $O_{kn}$ between closed eigenstates and open non-Hermitian resonance modes.
3) \textit{Spectral Mpemba Crossings:} 
We introduced a diagonal spectral relaxation measure $S(t,T_i)$ that cleanly isolates mode decay from off-diagonal interference. 
For a quartic double well, we demonstrated that $S(t,T_i)$ displays finite-time temperature crossings, where an initially hotter preparation relaxes faster than an initially colder one.
4) \textit{Spectral vs. State-Space Dynamics:} By computing the conditional trace distance $D_{\text{QS}}(t,T_i)$ relative to the asymptotic quasi-stationary state, we found no trace-distance crossing for the initial temperatures and parameter regime examined. 
This demonstrates that the anomaly is specifically a \emph{spectral Mpemba effect} rather than an overall state-space acceleration.

Our continuous-space framework successfully bridges real-space intuition regarding boundary-driven loss with the mathematical principles of non-Hermitian quantum mechanics. 
These results provide clear guidelines for future experimental setups, such as optical lattices with localized loss, dissipative photonic waveguides, or ultracold atomic systems, where spectral filter measures and boundary-engineered dissipation can be leveraged to control quantum thermal relaxation.

\section*{Acknowledgment}
We thank Y. Liu for stimulating discussion.
HH also thanks F. Nori, R. Hamazaki, N. Ohga, F. van Wijland, and T. Van Vu for fruitful discussions.
This work is partially supported by JSPS Kakenhi (Grant Nos.~JP24K06974, JP24K07193, JP25K01063, and JP26K06960).

\newpage
\appendix

\section{Spectral Representation of the Conditional Dynamics}
\label{app:conditional_spectral}
In this Appendix, we summarize the spectral representation of the subnormalized conditional density operator generated by the non-Hermitian effective Hamiltonian $\hat H_{\rm eff}$.
This representation clarifies the roles of the diagonal and off-diagonal resonance coefficients and their relation to the diagonal spectral relaxation measure introduced in the main text.

\subsection{Biorthogonal representation}
The right and left eigenstates of the non-Hermitian effective
Hamiltonian are defined by
\begin{align}
    \hat H_{\rm eff}|\phi_k\rangle
    &=
    \lambda_k|\phi_k\rangle,
    \qquad
    \lambda_k
    =
    \mathcal E_k-i\frac{\Gamma_k}{2},
    \label{eq:right_eigen_appA}
    \\
    \hat H_{\rm eff}^\dagger|\chi_k\rangle
    &=
    \lambda_k^*|\chi_k\rangle.
    \label{eq:left_eigen_appA}
\end{align}
Here, $\mathcal E_k=\operatorname{Re}\lambda_k$ is the resonance energy and $\Gamma_k=-2\operatorname{Im}\lambda_k\geq0$ is the corresponding decay rate.

We assume that $\hat H_{\rm eff}$ is diagonalizable and that the left and right eigenstates form a complete biorthogonal set,
\begin{equation}
    \langle\chi_j|\phi_k\rangle
    =
    \delta_{jk},
    \qquad
    \sum_k
    |\phi_k\rangle\langle\chi_k|
    =
    \hat{\mathbb I}.
    \label{eq:biortho_norm_appA}
\end{equation}

For the CAP Hamiltonian considered here, the coordinate-space representation is complex symmetric,
\begin{equation}
    \hat H_{\rm eff}^{T}
    =
    \hat H_{\rm eff}.
\end{equation}
The left and right eigenfunctions can therefore be chosen such that
\begin{equation}
    \chi_k(x)=\phi_k^*(x),
    \qquad
    \langle\chi_k|
    =
    \int_{-L}^{L}dx\,
    \phi_k(x)\langle x|,
    \label{eq:complex_symm_relation_appA}
\end{equation}
with the normalization
\begin{equation}
    \int_{-L}^{L}dx\,
    \phi_j(x)\phi_k(x)
    =
    \delta_{jk}.
    \label{eq:c_product_appA}
\end{equation}
Equation~\eqref{eq:c_product_appA} is a c-product orthogonality relation and must be distinguished from the ordinary Hilbert-space inner product
\begin{equation}
    \langle\phi_j|\phi_k\rangle
    =
    \int_{-L}^{L}dx\,
    \phi_j^*(x)\phi_k(x).
    \label{eq:ordinary_inner_appA}
\end{equation}

\subsection{Full spectral expansion of the conditional density operator}
The no-jump evolution operator is
\begin{equation}
    \hat U_{\rm eff}(t)
    =
    e^{-i\hat H_{\rm eff}t/\hbar}.
\end{equation}
Using the completeness relation in Eq.~\eqref{eq:biortho_norm_appA}, it can be written as
\begin{equation}
    \hat U_{\rm eff}(t)
    =
    \sum_k
    e^{-i\lambda_k t/\hbar}
    |\phi_k\rangle\langle\chi_k|.
    \label{eq:spectral_U_appA}
\end{equation}

The initial thermal state is
\begin{equation}
    \hat\rho_i(T_i)
    =
    \sum_n
    P_n(T_i)|n\rangle\langle n|,
    \qquad
    P_n(T_i)
    =
    \frac{
        e^{-E_n/(k_BT_i)}
    }{
        Z(T_i)
    }.
    \label{eq:initial_rho_appA}
\end{equation}
The subnormalized no-jump density operator evolves according to
\begin{equation}
    \hat\rho_c(t,T_i)
    =
    \hat U_{\rm eff}(t)
    \hat\rho_i(T_i)
    \hat U_{\rm eff}^\dagger(t).
    \label{eq:rho_t_evolved_appA}
\end{equation}

Substitution of Eq.~\eqref{eq:spectral_U_appA} into Eq.~\eqref{eq:rho_t_evolved_appA} yields
\begin{equation}
    \hat\rho_c(t,T_i)
    =
    \sum_{k,j}
    a_{k,j}(T_i)
    e^{-i(\lambda_k-\lambda_j^*)t/\hbar}
    |\phi_k\rangle\langle\phi_j|,
    \label{eq:rho_t_full_appA}
\end{equation}
where
\begin{equation}
    a_{k,j}(T_i)
    :=
    \langle\chi_k|
    \hat\rho_i(T_i)
    |\chi_j\rangle.
    \label{eq:akj_appA}
\end{equation}

Introducing
\begin{equation}
    O_{kn}
    :=
    \langle\chi_k|n\rangle
    =
    \int_{-L}^{L}
    dx\,
    \phi_k(x)\varphi_n(x),
    \qquad
    \varphi_n(x):=\langle x|n\rangle,
    \label{eq:Okn_appA}
\end{equation}
we obtain
\begin{equation}
    a_{k,j}(T_i)
    =
    \frac{1}{Z(T_i)}
    \sum_n
    e^{-E_n/(k_BT_i)}
    O_{kn}O_{jn}^*.
    \label{eq:akj_overlap_appA}
\end{equation}
In particular,
\begin{equation}
    a_k(T_i)
    :=
    a_{k,k}(T_i)
    =
    \frac{1}{Z(T_i)}
    \sum_n
    e^{-E_n/(k_BT_i)}
    |O_{kn}|^2
    \geq0.
    \label{eq:akk_appA}
\end{equation}

Equation~\eqref{eq:rho_t_full_appA} shows explicitly that the complete conditional density-matrix dynamics contains both diagonal and off-diagonal resonance contributions.  
The latter contain relative phase factors between different resonance energies and therefore encode cross-mode interference that is absent from the diagonal spectral relaxation measure.

For the reflection-symmetric potential and CAP considered in the main text, the resonance modes and the closed-system eigenstates have definite parity.  
Consequently,
\begin{equation}
    a_{k,j}(T_i)=0
\end{equation}
when $k$ and $j$ belong to opposite parity sectors, whereas same-parity off-diagonal coefficients generally remain finite.

\subsection{Normalization of the conditional state}
The trace of the subnormalized density operator gives the probability that the particle remains in the one-particle sector up to time $t$, which we denote by
\begin{equation}
    P_{\rm surv}(t,T_i)
    :=
    \mathrm{Tr}\hat\rho_c(t,T_i).
    \label{eq:Psurv_appA}
\end{equation}
Taking the trace of Eq.~\eqref{eq:rho_t_full_appA} gives
\begin{equation}
    P_{\rm surv}(t,T_i)
    =
    \sum_{k,j}
    a_{k,j}(T_i)
    e^{-i(\lambda_k-\lambda_j^*)t/\hbar}
    G_{jk},
    \label{eq:Psurv_exact_appA}
\end{equation}
where
\begin{equation}
    G_{jk}
    :=
    \langle\phi_j|\phi_k\rangle
    =
    \int_{-L}^{L}dx\,
    \phi_j^*(x)\phi_k(x)
    \label{eq:Gram_appA}
\end{equation}
is the ordinary Gram matrix of the right resonance eigenstates.

The normalized conditional state used in the trace-distance analysis is therefore
\begin{equation}
    \hat\rho_{\rm cond}(t,T_i)
    =
    \frac{
        \hat\rho_c(t,T_i)
    }{
        P_{\rm surv}(t,T_i)
    }.
    \label{eq:rho_cond_appA}
\end{equation}
Both the numerator and the normalization factor generally contain same-parity off-diagonal contributions.

It is important that the c-product orthogonality in Eq.~\eqref{eq:c_product_appA} does not imply
\begin{equation}
    G_{jk}=\delta_{jk}.
\end{equation}
Indeed, the resonance equations imply
\begin{equation}
    (\lambda_k-\lambda_j^*)
    \int_{-L}^{L}dx\,
    \phi_j^*(x)\phi_k(x)
    =
    -2i
    \int_{-L}^{L}dx\,
    \phi_j^*(x)W(x)\phi_k(x),
    \label{eq:ordinary_overlap_relation_appA}
\end{equation}
under the Dirichlet boundary conditions
$\phi_k(\pm L)=0$.
The right-hand side is generally nonzero, so that $\langle\phi_j|\phi_k\rangle$ need not vanish for $j\neq k$.

For the reflection-symmetric system, $G_{jk}$ vanishes when $j$ and $k$ have opposite parity.  
Same-parity off-diagonal terms, however, are not eliminated by symmetry and remain part of the full conditional-state dynamics.

\subsection{Relation to the diagonal spectral relaxation measure}
The spectral relaxation measure introduced in the main text is
\begin{equation}
     S(t,T_i)
    =
    \sum_k
    a_k(T_i)e^{-\Gamma_k t/\hbar}.
    \label{eq:Stilde_appA}
\end{equation}
This quantity should be distinguished from $P_{\rm surv}(t,T_i)$ in Eq.~\eqref{eq:Psurv_exact_appA}.
The latter is obtained from the trace of the full conditional density operator and generally contains both the Gram matrix $G_{jk}$ and off-diagonal coefficients $a_{k,j}$.

By contrast, $S(t,T_i)$ is intentionally constructed only from the diagonal spectral weights $a_k(T_i)$ and the corresponding decay rates $\Gamma_k$. 
It therefore isolates the decay-rate filtering of the thermal resonance spectrum:
\begin{equation}
    S(t,T_i)
    =
    \sum_k a_k(T_i)e^{-\Gamma_k t/\hbar},
\end{equation}
where $a_k(T_i)$ is the thermal spectral weight of the $k$-th resonance mode, and $e^{-\Gamma_k t/\hbar}$ describes its decay with the rate $\Gamma_k/\hbar$.

Thus, $S(t,T_i)$ is not obtained by approximating $P_{\rm surv}(t,T_i)$ and neglecting off-diagonal terms.
Rather, the two quantities characterize different aspects of the open-system dynamics: $P_{\rm surv}$ describes total particle retention, whereas $S$ is a diagonal spectral relaxation measure.  
The normalized state $\hat\rho_{\rm cond}(t,T_i)$ retains the full spectral information and is used separately in the trace-distance analysis of Sec.~\ref{sec:trace_distance}.

\section{Low-Temperature Bound on the Diagonal Coefficients
$a_k(T_i)$}
\label{app:low_temperature_ak}

In this Appendix, we examine the low-temperature behavior of the
diagonal spectral coefficients
\begin{equation}
    a_k(T_i):=a_{k,k}(T_i)
    =
    \frac{1}{Z(T_i)}
    \sum_{m\in\mathcal P_k}
    e^{-E_m/(k_BT_i)}
    |O_{km}|^2,
    \qquad
    k\neq0,
    \label{eq:ak_appendix_def}
\end{equation}
where $\mathcal P_k$ denotes the set of closed-system eigenstates
having the same parity as the $k$-th resonance mode, and
\begin{equation}
    O_{km}:=\langle\chi_k|m\rangle
    =
    \int_{-L}^{L}
    \phi_k(x)\varphi_m(x)\,dx.
    \label{eq:Okm_appendix}
\end{equation}
Here, $|m\rangle$ is an eigenstate of the closed Hamiltonian
$\hat H_0$, whereas $\langle\chi_k|$ is the left eigenstate of the
effective non-Hermitian Hamiltonian
\begin{equation}
    \hat H_{\mathrm{eff}}
    =
    \hat H_0-i\hat W.
    \label{eq:Heff_lowT_appendix}
\end{equation}
For the complex-symmetric operator considered here,
$\chi_k^*(x)=\phi_k(x)$, which leads to the second equality in
Eq.~\eqref{eq:Okm_appendix}.

\subsection{Exact low-temperature limit}

Let
\begin{equation}
    E_0<E_1<E_2<\cdots
\end{equation}
denote the eigenvalues of the closed Hamiltonian.
Factoring out the ground-state Boltzmann factor, the partition
function can be written as
\begin{equation}
    Z(T_i)
    =
    e^{-E_0/(k_BT_i)}
    \left[
        1+
        \sum_{m\geq1}
        e^{-(E_m-E_0)/(k_BT_i)}
    \right].
    \label{eq:partition_lowT_appendix}
\end{equation}

For an even resonance mode, Eq.~\eqref{eq:ak_appendix_def} becomes
\begin{equation}
    a_k(T_i)
    =
    \frac{
        |O_{k0}|^2+
        \displaystyle
        \sum_{\substack{m\geq2\\m\in\mathcal P_k}}
        e^{-(E_m-E_0)/(k_BT_i)}
        |O_{km}|^2
    }{
        1+
        \displaystyle
        \sum_{m\geq1}
        e^{-(E_m-E_0)/(k_BT_i)}
    }.
    \label{eq:ak_factored_appendix_even}
\end{equation}
Since $E_m-E_0>0$ for every $m\geq1$, all the excited-state
Boltzmann factors vanish as $T_i\to0^+$.
Therefore,
\begin{equation}
    \lim_{T_i\to0^+}a_k(T_i)
    =
    |O_{k0}|^2
    \qquad
    (k:\ {\rm even}).
    \label{eq:ak_lowT_appendix_even}
\end{equation}

Thus, for an even resonance mode with $k\neq0$, the low-temperature
limit is determined entirely by the projection of the closed-system
ground state onto the $k$-th left eigenmode of
$\hat H_{\mathrm{eff}}$.

In general, $O_{k0}$ does not vanish identically for even $k$.
Although the closed-system eigenstates satisfy
\begin{equation}
    \langle k|0\rangle=0,
    \qquad
    k\neq0,
\end{equation}
the states $\langle\chi_k|$ and $|0\rangle$ are eigenstates of
different operators. Consequently, the orthogonality theorem for
the Hermitian operator $\hat H_0$ cannot be applied to
$\langle\chi_k|0\rangle$.

For an odd resonance mode, on the other hand, reflection symmetry
implies
\begin{equation}
    O_{k0}
    =
    \langle\chi_k|0\rangle
    =
    0,
    \qquad
    k:\ {\rm odd},
    \label{eq:Ok0_odd}
\end{equation}
because the closed-system ground state is even.
Consequently,
\begin{equation}
    \lim_{T_i\to0^+}a_k(T_i)
    =
    0,
    \qquad
    k:\ {\rm odd}.
    \label{eq:ak_lowT_appendix_odd}
\end{equation}

Combining the two cases, the exact low-temperature limit can be
written compactly as
\begin{equation}
    \lim_{T_i\to0^+}a_k(T_i)
    =
    |O_{k0}|^2
    \qquad
    (k\neq0),
    \label{eq:ak_lowT_appendix}
\end{equation}
where $O_{k0}=0$ identically for odd $k$.

\subsection{Exact identity in terms of the absorbing potential}

The left eigenstate satisfies
\begin{equation}
    \langle\chi_k|\hat H_{\mathrm{eff}}
    =
    \lambda_k\langle\chi_k|,
    \qquad
    \lambda_k
    =
    E_k^{(\mathrm{eff})}
    -i\frac{\Gamma_k}{2},
    \label{eq:left_eigen_appendix}
\end{equation}
while the closed-system ground state satisfies
\begin{equation}
    \hat H_0|0\rangle
    =
    E_0|0\rangle.
    \label{eq:closed_ground_appendix}
\end{equation}

Taking the matrix element of Eq.~\eqref{eq:left_eigen_appendix}
with $|0\rangle$ and using
$\hat H_{\mathrm{eff}}=\hat H_0-i\hat W$, we obtain
\begin{align}
    \lambda_k\langle\chi_k|0\rangle
    &=
    \langle\chi_k|
    \left(
        \hat H_0-i\hat W
    \right)
    |0\rangle
    \nonumber\\
    &=
    E_0\langle\chi_k|0\rangle
    -
    i\langle\chi_k|\hat W|0\rangle.
    \label{eq:Ok0_identity_derivation}
\end{align}
Provided that $\lambda_k\neq E_0$, this gives the exact identity
\begin{equation}
    O_{k0}
    =
    -
    \frac{
        i\langle\chi_k|\hat W|0\rangle
    }{
        \lambda_k-E_0
    }.
    \label{eq:Ok0_exact_identity}
\end{equation}
Accordingly,
\begin{equation}
    |O_{k0}|
    =
    \frac{
        |\langle\chi_k|\hat W|0\rangle|
    }{
        |\lambda_k-E_0|
    }.
    \label{eq:Ok0_exact_modulus}
\end{equation}

Equation~\eqref{eq:Ok0_exact_identity} shows that, for even
$k\neq0$, the nonzero ground-state overlap is generated by the
absorbing potential.
In particular,
\begin{equation}
    O_{k0}=0
\end{equation}
holds when
\begin{equation}
    \langle\chi_k|\hat W|0\rangle=0.
    \label{eq:Ok0_zero_condition}
\end{equation}

For odd $k$, this condition is automatically satisfied in the
reflection-symmetric system considered here.  Indeed, when
$W(x)=W(-x)$, the integrand in
$\langle\chi_k|\hat W|0\rangle$ is odd, and hence
\begin{equation}
    \langle\chi_k|\hat W|0\rangle=0,
    \qquad
    k:\ {\rm odd}.
\end{equation}
Thus, Eq.~\eqref{eq:Ok0_exact_identity} is fully consistent with the
parity relation \eqref{eq:Ok0_odd}.

\subsection{Upper bound in the absorbing region}

Let
\begin{equation}
    \Omega_{\mathrm{CAP}}
    :=
    \left\{
        x\in[-L,L]\;|\;W(x)\neq0
    \right\}
    \label{eq:CAP_region}
\end{equation}
denote the support of the complex absorbing potential.
In coordinate representation,
\begin{equation}
    \langle\chi_k|\hat W|0\rangle
    =
    \int_{\Omega_{\mathrm{CAP}}}
    \phi_k(x)W(x)\varphi_0(x)\,dx.
    \label{eq:CAP_matrix_element}
\end{equation}
Taking the absolute value and applying the Cauchy--Schwarz
inequality gives
\begin{align}
    |\langle\chi_k|\hat W|0\rangle|
    &\leq
    \left[
        \int_{\Omega_{\mathrm{CAP}}}
        |\phi_k(x)|^2\,dx
    \right]^{1/2}
    \left[
        \int_{\Omega_{\mathrm{CAP}}}
        W(x)^2|\varphi_0(x)|^2\,dx
    \right]^{1/2}.
    \label{eq:CAP_CS_bound}
\end{align}

Introducing the restricted norm
\begin{equation}
    \|f\|_{\Omega_{\mathrm{CAP}}}^2
    :=
    \int_{\Omega_{\mathrm{CAP}}}
    |f(x)|^2\,dx
    \label{eq:restricted_norm}
\end{equation}
and
\begin{equation}
    W_{\max}
    :=
    \sup_{x\in\Omega_{\mathrm{CAP}}}W(x),
\end{equation}
we obtain
\begin{equation}
    |\langle\chi_k|\hat W|0\rangle|
    \leq
    W_{\max}
    \|\phi_k\|_{\Omega_{\mathrm{CAP}}}
    \|\varphi_0\|_{\Omega_{\mathrm{CAP}}}.
    \label{eq:CAP_simple_bound}
\end{equation}

Combining Eqs.~\eqref{eq:Ok0_exact_modulus} and
\eqref{eq:CAP_simple_bound}, we find
\begin{equation}
    |O_{k0}|
    \leq
    \frac{
        W_{\max}
        \|\phi_k\|_{\Omega_{\mathrm{CAP}}}
        \|\varphi_0\|_{\Omega_{\mathrm{CAP}}}
    }{
        |\lambda_k-E_0|
    }.
    \label{eq:Ok0_upper_bound}
\end{equation}
Therefore, the exact low-temperature value satisfies
\begin{equation}
    \lim_{T_i\to0^+}a_k(T_i)
    =
    |O_{k0}|^2
    \leq
    \frac{
        W_{\max}^2
        \|\phi_k\|_{\Omega_{\mathrm{CAP}}}^2
        \|\varphi_0\|_{\Omega_{\mathrm{CAP}}}^2
    }{
        |\lambda_k-E_0|^2
    }.
    \label{eq:ak_lowT_upper_bound}
\end{equation}

For odd $k$, the left-hand side vanishes exactly by parity.
For even $k\neq0$, the bound gives a quantitative criterion for
the smallness of the low-temperature coefficient.
It is small when the closed-system ground state has only a small
tail in the absorbing region, provided that the open-system mode
norm in this region remains bounded and $\lambda_k$ is sufficiently
separated from $E_0$.

\subsection{Exponential suppression for a remote absorber}
Suppose that the absorbing region lies beyond the outer classical turning point $x_{\mathrm{tp}}$ of the closed-system ground state.
Under the usual semiclassical localization assumptions, the ground-state wave function in this classically forbidden region satisfies an estimate of the form
\begin{equation}
    \|\varphi_0\|_{\Omega_{\mathrm{CAP}}}
    \leq
    C_0
    \exp\left(
        -\frac{\mathcal A_0}{\hbar}
    \right),
    \label{eq:ground_Agmon_bound}
\end{equation}
where $C_0$ is independent of the distance between the core and the absorbing region, and
\begin{equation}
    \mathcal A_0
    :=
    \int_{x_{\mathrm{tp}}}^{x_{\mathrm{cap}}}
    \sqrt{
        2M\left[V(x)-E_0\right]
    }\,dx.
    \label{eq:ground_action}
\end{equation}
Here, $M$ denotes the particle mass and $x_{\mathrm{cap}}$ is the position at which the CAP begins.

Equation~\eqref{eq:Ok0_upper_bound} then implies
\begin{equation}
    |O_{k0}|
    \leq
    \frac{
        C_0W_{\max}
        \|\phi_k\|_{\Omega_{\mathrm{CAP}}}
    }{
        |\lambda_k-E_0|
    }
    \exp\left(
        -\frac{\mathcal A_0}{\hbar}
    \right).
    \label{eq:Ok0_exponential_bound}
\end{equation}
Consequently,
\begin{equation}
    \lim_{T_i\to0^+}a_k(T_i)
    \leq
    \frac{
        C_0^2W_{\max}^2
        \|\phi_k\|_{\Omega_{\mathrm{CAP}}}^2
    }{
        |\lambda_k-E_0|^2
    }
    \exp\left(
        -\frac{2\mathcal A_0}{\hbar}
    \right).
    \label{eq:ak_exponential_bound}
\end{equation}

For even $k\neq0$, this shows that the low-temperature coefficient is exponentially suppressed when the absorbing layer is sufficiently separated from the localized ground state.
This provides a mathematical justification for treating $a_k(0^+)$ as numerically negligible in the corresponding parameter regime, without assuming that it vanishes exactly.

For odd $k$, the stronger result
\begin{equation}
    a_k(0^+)=0
\end{equation}
holds exactly by parity.

\subsection{Weak-CAP expansion}
The same conclusion can be understood perturbatively.
For a weak absorbing potential and an isolated $k$-th excited state, non-Hermitian perturbation theory gives
\begin{equation}
    |\phi_k\rangle
    =
    |k\rangle
    -
    i\sum_{\ell\neq k}
    \frac{
        \langle\ell|\hat W|k\rangle
    }{
        E_k-E_\ell
    }
    |\ell\rangle
    +
    \mathcal O(\|\hat W\|^2).
    \label{eq:phik_weak_CAP}
\end{equation}
Since
\begin{equation}
    \langle0|k\rangle=0,
    \qquad
    k\neq0,
\end{equation}
the leading contribution to the ground-state overlap is
\begin{equation}
    O_{k0}
    =
    -i
    \frac{
        \langle0|\hat W|k\rangle
    }{
        E_k-E_0
    }
    +
    \mathcal O(\|\hat W\|^2).
    \label{eq:Ok0_weak_CAP}
\end{equation}
It follows that
\begin{equation}
    |O_{k0}|^2
    =
    \frac{
        |\langle0|\hat W|k\rangle|^2
    }{
        (E_k-E_0)^2
    }
    +
    \mathcal O(\|\hat W\|^3).
    \label{eq:Ok0_square_weak_CAP}
\end{equation}

For an even resonance mode with $k\neq0$,
$O_{k0}$ is therefore of first order in the CAP strength, whereas the low-temperature coefficient
\begin{equation}
    a_k(0^+)=|O_{k0}|^2
\end{equation}
is of second order.

For odd $k$, reflection symmetry of $\hat W$ gives
\begin{equation}
    \langle0|\hat W|k\rangle=0,
\end{equation}
and the exact parity relation
\begin{equation}
    O_{k0}=0
\end{equation}
holds to all orders rather than only perturbatively.

The perturbative estimate for even $k$ requires that the CAP be sufficiently weak, that the relevant eigenvalue be isolated, and that the system not be close to an exceptional point.
The exact bound~\eqref{eq:ak_lowT_upper_bound}, by contrast, does not rely on a weak-CAP expansion, although it can become uninformative if $\|\phi_k\|_{\Omega_{\mathrm{CAP}}}$ is strongly enhanced by non-normality.

\subsection{Conclusion}
For every nontrivial resonance mode $k\neq0$, the exact low-temperature limit is
\begin{equation}
    \lim_{T_i\to0^+}a_k(T_i)
    =
    |O_{k0}|^2.
\end{equation}
For odd $k$, reflection symmetry gives
\begin{equation}
    |O_{k0}|^2=0
\end{equation}
exactly.

For even $k\neq0$, the overlap $O_{k0}$ is controlled by the CAP matrix element $\langle\chi_k|\hat W|0\rangle$ through Eq.~\eqref{eq:Ok0_exact_identity}.
When the absorbing layer is located sufficiently far from the double-well core, the exponentially small ground-state amplitude in the absorbing region makes $|O_{k0}|^2$ correspondingly small.
Therefore, for even nontrivial resonance modes, the vanishing low-temperature value used in an idealized description should be
understood as the asymptotic approximation
\begin{equation}
    |O_{k0}|^2\ll1,
    \qquad
    k\neq0,\quad k:\ {\rm even},
\end{equation}
rather than as an exact orthogonality relation.

These results explain why the nontrivial diagonal spectral weights $a_k(T_i)$ can be weak or absent at low temperature and become thermally activated at finite temperature.  
Their different temperature dependences, together with the mode-dependent decay rates $\Gamma_k$, determine the spectral relaxation measure
\begin{equation}
     S(t,T_i)
    =
    \sum_k
    a_k(T_i)e^{-\Gamma_k t/\hbar}.
\end{equation}

\section{Details of the numerical implementation}
\label{sec:numerical}
In this Appendix, we describe the numerical procedure used to evaluate the resonance eigenstates, overlap coefficients, spectral relaxation measures, and conditional-state quantities presented in the main text.
Throughout the numerical calculations, we use dimensionless units with
\begin{equation}
    m=\hbar=k_B=1.
\end{equation}

\subsection{Spatial discretization}

The one-dimensional coordinate space is discretized on a uniform grid,
\begin{equation}
    x_i=-L+i\Delta x,
    \qquad
    i=0,\ldots,N-1,
\end{equation}
with
\begin{equation}
    \Delta x=\frac{2L}{N-1}.
\end{equation}
The grid includes the two boundary points at $x=\pm L$.
The Dirichlet boundary conditions are imposed by eliminating these endpoints, so that the resulting Hamiltonian matrices have dimension
\begin{equation}
    N_{\rm int}=N-2.
\end{equation}
Throughout the main calculations, we use $N=1200$, corresponding to a matrix dimension $N_{\rm int}=1198$.
The numerical convergence of this choice is examined below.

The kinetic-energy operator is approximated by the second-order central finite-difference formula,
\begin{equation}
    -\frac{1}{2}\frac{d^2\psi(x)}{dx^2}
    \approx
    -\frac{1}{2}
    \frac{
        \psi(x+\Delta x)-2\psi(x)+\psi(x-\Delta x)
    }{\Delta x^2}.
\end{equation}
Consequently, both the Hermitian Hamiltonian $H_0$ and the effective non-Hermitian Hamiltonian $H_{\rm eff}$ are represented by finite-dimensional matrices.

The complex absorbing potential is taken to be quadratic, corresponding to the exponent $n=2$. Thus, in the absorbing region, the CAP increases quadratically with the distance from its onset position.

\subsection{Hermitian and non-Hermitian eigenvalue problems}
The discretized Hermitian Hamiltonian is diagonalized by solving
\begin{equation}
    H_0\bm{\psi}_n
    =
    E_n\bm{\psi}_n,
\end{equation}
where $\bm{\psi}_n$ denotes the discretized eigenfunction associated with the eigenvalue $E_n$.
Since $H_0$ is Hermitian, its eigenvalues are real and its eigenvectors form an orthonormal basis.
The eigenvalue problem is solved numerically using a standard dense Hermitian eigensolver.

The effective Hamiltonian,
\begin{equation}
    H_{\rm eff}
    =
    H_0-iW,
\end{equation}
is non-Hermitian because of the complex absorbing potential.
Its right eigenvectors are obtained from
\begin{equation}
    H_{\rm eff}\bm{\phi}_k
    =
    \lambda_k\bm{\phi}_k,
\end{equation}
where the complex eigenvalues are written as
\begin{equation}
    \lambda_k
    =
    \mathcal{E}_k-\frac{i}{2}\Gamma_k.
\end{equation}
Here, $\mathcal{E}_k=\operatorname{Re}\lambda_k$ is the resonance energy and $\Gamma_k=-2\operatorname{Im}\lambda_k$ is the corresponding resonance width.
We use $\mathcal{E}_k$ here to distinguish the resonance energies from the closed-system eigenvalues $E_n$.

The left eigenvectors are obtained by solving the adjoint eigenvalue problem,
\begin{equation}
    H_{\rm eff}^{\dagger}\bm{\chi}_k
    =
    \lambda_k^*\bm{\chi}_k,
\end{equation}
where $\lambda_k^*$ denotes the complex conjugate of $\lambda_k$.
After diagonalization, the left and right eigenvectors are normalized so as to satisfy the biorthogonality relation
\begin{equation}
    \bm{\chi}_j^{\dagger}\bm{\phi}_k
    =
    \delta_{jk}.
\end{equation}
This normalization is used throughout the numerical calculations.

All $N_{\rm int}=N-2$ eigenstates of the discretized closed-system Hamiltonian are included in the construction of the Gibbs state.
By contrast, the spectral relaxation measure is evaluated by retaining the first $M=80$ resonance states of the effective non-Hermitian Hamiltonian, ordered by increasing real part of the resonance
eigenvalue.

\subsection{Continuum and finite-grid high-temperature limits}

The high-temperature limit must be distinguished between the continuum theory and its finite-dimensional discretization.
In the continuum on a finite spatial interval, the number of closed-system energy levels is infinite. 
Consequently,
\begin{equation}
    Z(T_i)
    =
    \sum_n e^{-E_n/T_i}
    \longrightarrow\infty
    \qquad
    (T_i\to\infty).
\end{equation}
For a fixed resonance state $k$, the completeness of the closed-system eigenstates gives a finite numerator,
\begin{equation}
    \sum_n
    \left|
        \langle\chi_k|n\rangle
    \right|^2
    =
    \langle\chi_k|\chi_k\rangle.
\end{equation}
It therefore follows that
\begin{equation}
    a_k(T_i)
    \longrightarrow 0
    \qquad
    (T_i\to\infty)
\end{equation}
in the continuum limit.

For a fixed spatial grid, however, the Hilbert-space dimension is finite and equal to $N_{\rm int}=N-2$.
The Gibbs probabilities then approach the uniform distribution,
\begin{equation}
    P_n(T_i)
    \longrightarrow
    \frac{1}{N_{\rm int}}
    \qquad
    (T_i\to\infty),
\end{equation}
and hence
\begin{equation}
    a_k(T_i)
    \longrightarrow
    \frac{
        \langle\chi_k|\chi_k\rangle
    }{N_{\rm int}}.
\end{equation}
Thus, the small nonzero high-temperature limit observed for a fixed grid is a finite-dimensional effect and vanishes as the continuum
limit is approached.

\subsection{Numerical convergence}
\label{sec:numerical-convergence}
To examine numerical convergence, we introduce the notation $S_{N,M}(t,T_i)$.
The first subscript $N$ denotes the number of spatial grid points, whereas the second subscript $M$ denotes the number of resonance states retained in the spectral sum.
The spectral relaxation measure evaluated with these numerical parameters is therefore
\begin{equation}
    S_{N,M}(t,T_i)
    =
    \sum_{k=0}^{M-1}
    a_k^{(N)}(T_i)
    e^{-\Gamma_k^{(N)}t}.
\end{equation}
Here, the superscript $(N)$ indicates that both the spectral weight and the resonance width are calculated using an $N$-point spatial grid.
The notation $S_{N,M}$ is introduced only to describe the numerical convergence; the subscripts are omitted elsewhere in the manuscript.

\begin{figure}[htbp]
    \centering
    \includegraphics[width=\textwidth]{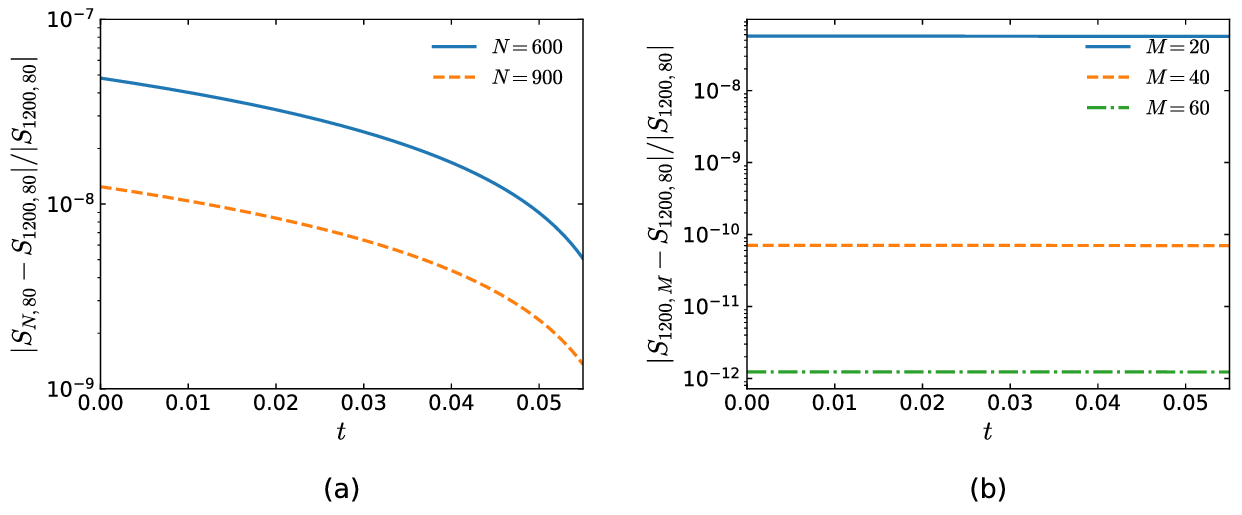}
    \caption{
    Numerical convergence of the spectral relaxation measure at $T_i=2$.
    Here, $S_{N,M}(t,T_i)$ denotes the result calculated using $N$ spatial grid points and $M$ retained resonance states.
    (a) Relative deviations for $N=600$ and $900$ at fixed $M=80$, using $S_{1200,80}$ as the reference.
    (b) Relative deviations for $M=20$, $40$, and $60$ at fixed $N=1200$, again using $S_{1200,80}$ as the reference.
    In both panels, all other numerical and physical parameters are held fixed.
    }
    \label{fig:convergence}
\end{figure}

Figure~\ref{fig:convergence} shows the relative deviations from the reference calculation $S_{1200,80}(t,T_i)$ for the representative initial temperature $T_i=2$.
Figure~\ref{fig:convergence}(a) examines spatial-grid convergence by varying $N$ while fixing $M=80$.
The relative deviation remains below approximately $5\times10^{-8}$ even for $N=600$, and below approximately $1.3\times10^{-8}$ for $N=900$.

Figure~\ref{fig:convergence}(b) examines convergence with respect to the number of retained resonance states by varying $M$ while fixing $N=1200$.
The relative deviation remains below approximately $6\times10^{-8}$ for $M=20$, $10^{-10}$ for $M=40$, and $2\times10^{-12}$ for $M=60$.
These results confirm that the numerical choice $N=1200$ and $M=80$ is sufficiently converged for the spectral relaxation calculations reported in the main text.

\end{document}